\documentclass[reprint,amssymb,amsmath,aip,cha]{revtex4-2}
\usepackage{graphicx}
\usepackage[caption = false]{subfig}
\usepackage{color}
\usepackage{epsfig}
\usepackage{ifpdf}
\usepackage{bm}
\usepackage[utf8]{inputenc}
\usepackage{microtype}
\usepackage[colorlinks=true,linkcolor=blue]{hyperref}%
\expandafter\ifx\csname package@font\endcsname\relax\else
 \expandafter\expandafter
 \expandafter\usepackage
 \expandafter\expandafter
 \expandafter{\csname package@font\endcsname}% 
\fi
\usepackage{cleveref}
\begin{document}
\title{A REVIEW ON THE VORTEX AND COHERENT STRUCTURES IN DUSTY PLASMA MEDIUM}
\author{Mangilal Choudhary}
\email{mchoudhary@physics.du.ac.in}
\affiliation{Department of Physics and Astrophysics, University of Delhi, Delhi, India}
%\author{S. Mukherjee}
%\affiliation{Institute for Plasma Research, Bhat, Gandhinagar, 382428, India}
%\date{6-01-2024}
%
\begin{abstract}
Dusty plasma is an admixture of electrons, ions, and massive charged solid particles of sub-micron to micron sized in the background of neutral gas. The dust grain medium exhibits fluid (liquid) as well as solid-like characteristics at different background plasma conditions. It supports various linear and non-linear dynamical structures because of the external perturbation and internal instabilities. The vortical or coherent structure in the dusty plasma medium is a kind of self-sustained dynamical structure that is formed either by instabilities or external forcing. In this review article, the author discusses the past theoretical, experimental, and computational investigations on vortical and coherent structures in unmagnetized as well as in magnetized dusty plasma. The possible mechanisms to form vortices in dust grain medium are discussed in detail. The studies on evolution of vortices and their correlation with turbulence are also reviewed.
\end{abstract}
\maketitle
\section{Introduction}
Adding sub-micron to micron-sized solid dust particles to the plasma state of matter makes it more complex to study. In the plasma state having the neutral gas background, these solid particles (\textit{nm} to $\mu m$) undergo various charging processes and acquire negative or positive charges. In the laboratory plasma, dust particles are normally negatively charged \cite{chargingbarken,goreedustcharging1,daw2}. But, in afterglow conditions they acquire positive charges\cite{neerajpositivechargepop,afterglowpositivecharges2,chaubey_2023,afterglowpositivecharge1}. These charged dust particles experience finite gravitational force which leads them to settle down on the wall of the experimental device. A strong electric field using various discharge configurations \cite{pintudawdcconfiguration,mangilalpopdcconfiguration,linidustycrystal2,rfdischargeconfiguration,mangirsiexpsystem} opposite to gravity can hold dust grains for a longer time in the plasma medium. The confined dust grains in the electrostatic potential well created by using different discharge configurations exhibit a collective response under the action of internal or external electromagnetic perturbations. In the last 30 years, a broad spectrum of theoretical, experimental, and computational research work to explore the physics of dusty plasma medium which is an admixture of electrons, ions, and charged solid particles has been performed. The collective response of dusty plasma under electromagnetic perturbation is observed in the form of dust acoustic waves \cite{ddw1,ddw2,ddw3,dawmerlino,pintudawdcconfiguration,mangilalpopdcconfiguration,mangidawmagneticfield}, dust lattice waves \cite{dlw1,dlw2}, dust voids \cite{void,vorticesmicrogravity1999}, dusty plasma crystals \cite{linidustycrystal2,neerajcrystalccp1,surabhimagnetized2017}, rotational and vortex motion \cite{agarwalrotation,Satodustrotationinmagneticfield,Chaidustyicevortices2016,Bailung2020vortex,vaulinavortices2004,mangicppreview2023}, ring structures \cite{ringstructurevikram2023}, Mach cones \cite{machcones1999,machconespintu2017} etc.\\
\newline
%%%%%%%%%%%%%%%%%
A large amount of charge on massive solid particles makes dust grain medium unique compared to conventional two-component plasma. The dynamics of charged dust grains can be tracked easily because of their low frequency ($\sim$ 1 to 100 Hz) response \cite{dawmerlino,mangicppreview2023}. Moreover, the dust grain medium exhibits liquid and solid-like characteristics because of the higher average potential energy of the dust component compared to the kinetic energy \cite{phasetransitiondusty2008,mgdustycrystaltransition}. The fluid (liquid) or visco-elastic nature of the dust plasma system strongly depends on the coulomb coupling strength among charged dust grains. Therefore, the study of vortex and coherent structures similar to hydrodynamic fluid has been an active research topic in dusty plasma for the last 30 years. The study of the small-scale or the large-scale vortex structures in unmagnetized or magnetized dusty plasma helps to understand the heating of dust grain medium, heat transport mechanism, turbulence state of medium, diffusion and transport processes, mixing of different densities fluids, etc. The unique features of the dust plasma system could help in exploring the development and evolution of sheared flow, understanding the boundary values problems, formation of vortex structures, etc., in the colloidal systems\cite{colloidalvortices2021} and complex hydrodynamic fluids\cite{complexfluidvortices1,complexturbulentflowvortices1} at the molecular level. The analytical/computational/experimental study of localized potential structures or vortices in the liquid-like state of dusty plasma due to driven instabilities/turbulence may also play a significant role in understanding the development and evolution of the vortices in the astrophysical fluids or planetary systems or solar system such as Venus vortices\cite{venuspolevortices}, Jupiter vortices\cite{jupitorvortices}, Saturn's hexagon\cite{saturnhexagonvortices2018}, stratospheric vortices\cite{Stratosphericearthvortices}, polar vortices\cite{polarvorticesplant1}, tropical cyclones\cite{hurricanes2024}), Solar convection cells\cite{solarconvectionvortices}, Accretion discs\cite{acceretiondiskastro1981}, Spiral galaxies\cite{sprialgalaxy}) etc. by considering it as a model system.

%%%%%%%%%%%%%%%%%%%%%%
 In the last 30 years, extensive research work (theoretical, computational, and experimental) on the vortical and coherent structures in dusty plasma has been performed by different global research groups. It has been claimed that dust grain medium (a model system) has great potential to explore the dynamics of vortices observed in other physical systems at the kinetic level. However, there is a lack of a review article that complies with most of the research results on the vortices and coherent structures in dusty plasmas. Therefore, it is necessary to prepare a systematic review of the vortical and coherent structures in dusty plasma to benefit the research community. Given this objective, a review report that explores the basic physics of vortex formation and past research on the vortical structures in dust-plasma systems is prepared.\\
 \newline
%%%%%%%%%%%%%%%%%%%%%
The review article is organized as: Section \ref{sec:secII} deals with the mechanism for vortex formation and the role of boundary conditions on the formation and stability of vortices in dust grain medium. The results of vortex and coherent structures in unmagnetized dusty plasma are presented in Section \ref{sec:secIII}. Section \ref{sec:secIV} discusses the vortices in magnetized dusty plasma. The evolution and stability of dust vortices are discussed in Section \ref{sec:secV}. The connection between vortices and turbulence is reviewed in Section \ref{sec:secVI}. Concluding remarks along with future perspectives on the study of vortex motion in dusty plasma is given in Section \ref{sec:secVII}.
%%%%%%%%%%%%%%%%%%%%%
%%%%%%%%%%%
\section{Formation of the dust vortices}  \label{sec:secII}
This section deals with the formation of vortices due to the rotational motion of dust grains about a common center-line and the coherent structures, which are organized patterns that emerge due to the collective response of dust grain medium and persist in the background flow for an extended period. The dust dynamics (rotation/transport) is understood by exploring the forces acting upon charged dust grains in the plasma medium, possible free energy sources to drive the rotational/vortex motion in unmagnetized (magnetized) dusty plasma, and the role of the boundary conditions on the flow characteristics of dust grain medium in the respective subsections.  
\subsection{Forces acting upon dust particles}
The dynamics of charged dust particles are governed by the various forces acting on them in the plasma medium. A brief overview of these dust forces is given. 
\begin{itemize}
    \item [i.] \textbf{Gravitation force:}
The dust grains have finite mass; therefore, they will experience the force due to earth's gravity. For a spherical dust grain of mass $M_d$ , the gravitation force $F_g$ is \cite{shukladustybook,dustforcesnitter1996}
\begin{equation}
\vec{F_g} = M_d \vec{g} = \frac{4}{3} \pi r^3_d \rho_d \vec{g} ,
\end{equation}
 where $g$ is the acceleration due to gravity, $r_d$ is the radius of the dust grain and $\rho_d$ is the mass density of the dust grain.
 \item [ii.] \textbf{Electrostatic force:}
 The dust grains undergo various charging mechanisms in the plasma background and acquire a large amount of charge ($Q_d \sim 10^2 - 10^5 e^{-}$) on their surface. In laboratory discharges, the charged dust particles experience an electrostatic force $F_E$ due to the electric field $E$. The eclectic force is \cite{shukladustybook,dustforcesnitter1996}
\begin{equation}
\vec{F_E} = Q_d \vec{E} = \pm e Z_d \vec{E} ,
\end{equation} 
 where $Q_d = \pm e Z_d$ is the dust charge.
 \item [iii.] \textbf{Ion drag force:}
The dust grains are confined in the electrostatic potential well. The streaming ions in the electric field exert a force on the dust grains in two ways. The ions transfer momentum to a dust particle through direct impacts ($F_{ic}$). Secondly, the ions transfer momentum through coulomb collisions with the charged dust particles ($F_{io}$). The formulations of both forces \cite{dustforcesbarnes1992,dustforcesnitter1996,dustforcesshukla2002} under the approximation $\lambda_D/l_i <<1$, where $\lambda_D$ is the dust Debye (screening) length and $l_i$ is ion mean free path are given as- 
\begin{equation}
\centering
F_{ic} = n_i v_s m_i v_i \pi b^2_c,
\end{equation}
where $n_i$ is plasma density and $m_i$ is ion mass, $v_s$ the mean speed, $b_c$ is the collection impact parameter. 
 \begin{equation}
 \centering
 F_{io} = n_i v_s m_i v_i 4 \pi b^2_{\pi/2} \Gamma,
 \end{equation}
 where $b^2_{\pi/2}$ is the impact parameter whose asymptotic angle is $\pi/2$ and $\Gamma$ is the coulomb logarithm integrated over the interval from $b_c$ to $\lambda_D$.\\
 So the net ion drag force acting on charged grain is $F_i = F_{ic} + F_{io}$. Since $\vec{F_I}$ is acting in the direction of E-field, it can be written in terms of local E-field vector $\vec{F_I} =  F_i \hat{E}$. For the collisional dusty plasma, the updated formulation to estimate the ion drag force is required \cite{kharpakiondragfroce,iondragforcecolisonal}.  
%%%%%%%%%%%%%%
 \item [iv.] \textbf{Neutral drag force:}
 The neutral drag force is a kind of resistance experienced by the dust grains if they have the relative velocity or motion with respect to the background neutral gas. For the dust particles of size ($r_d$) smaller than the collision mean free path ($\lambda_{mpf})$, i.e. $r_d << \lambda_{mpf}$ and velocities much smaller than the thermal velocity of the gas, i.e. $v_d << v_{tn}$, the neutral drag force experienced by the dust particle is estimated using the Epstein's expression \cite{epsteinneutraldrag}
 \begin{equation}
 \centering
 \vec{F_n} = -M_d \nu_{dn} \vec{v_d},
 \end{equation}
 where $\nu_{dn}$ is the dust--neutral friction frequency and $v_d$ is the velocity of the dust particle relative to that of the neutral gas atoms.
 \item [v.] \textbf{Thermophoretic force:}
 The thermophoretic force arises due to a temperature gradient in the background neutral gas. The gas atoms present in the high-temperature zone exert more momentum on the confined dust particles than those present in the lower-temperature zone. Consequently, a thermophoretic force is established opposite to the temperature gradient. The magnitude of this force is \cite{shukladustybook} written as-
\begin{equation}
 \vec{F_{th}} = - \frac{32}{15} \frac{r^2_d}{v_{th,n}} \left(1 + \frac{5 \pi}{32} (1 - \alpha)\right)\kappa_T \nabla T_n,
\end{equation}
 where $\kappa_T$ is the translation thermal conductivity of the gas and $T_n$ is the neutral gas temperature. The value of $\alpha$ is $\approx $ 1 for dust particles and neutral gas atoms having temperatures below 500 K.
 \end{itemize}
%%%%%%%%%%%%%%%%%%%%%
%%%%%%%%%%%%%%%%%%%%%%%%%%%%%
\subsection{Physics of the driven vortices}
The observed vortices/coherent structures in dusty plasma can be categorized as externally driven and instability-driven vortices. In the externally driven vortices, dust grains behave as trace particles in the background of plasma. Since the dynamics of dust grain medium are associated with the background plasma species (electrons, ions, and neutrals), any internal or external electromagnetic perturbation alters the dynamics of ambient plasma species and sets the dust cloud into rotational motion.\\
In the fluid description of dusty plasma, the transport/motion of charged dust particles in the background of ambient plasma is understood by solving the equation of motion of dust particles under the action of driving and friction forces. The fluid equation of motion for dust grain medium which is assumed to be incompressible in nature is \cite{modelingvorticesmicrogravity1,modeingvorticesmicrogravity3}
\begin{equation}
M_d n_d \left(\frac{d\vec{v_d}}{dt}\right) = n_d (\vec{F_g} + \vec{F_E} + \vec{F_i} + \vec{F_n} + \vec{F_{th}}) -\nabla P_E + \eta \nabla^2 \vec{v_d}
\end{equation}
where $\vec{F_g}, \vec{F_E}, \vec{F_i}, \vec{F_n}$ and $ \vec{F_{th}}$ are the gravitational force, electric force, ion drag force, neutral drag force, and thermophoretic force acting on a single dust grain. $n_d$ is the dust density, $\vec{v_d}$ is the dust fluid element velocity, $\nabla P_E$ is the pressure force arises due to the dust-dust interactions and confinement of dust particles in an electrostatic potential well. $\eta$ is the kinetic viscosity of dust grain medium, and the whole term ($\eta \nabla^2 \vec{v_d}$) represents a kind of viscous force acting between layers of flowing dust grains. If dusty plasma experiments are performed at higher pressure (P $>$ 15 Pa), then the contribution of viscous friction can be ignored compared to the friction due to background neutral gas. In normal RF discharges and low-power DC discharges, the magnitude of the thermophoretic force acting on dust grains is a few orders lower than other dominant forces. Therefore, the contribution of thermophoresis force can also be dropped in the equation of motion (Eq. 7)    
\\
The vortex motion of dust grains in a 2-dimension (2D) plane can be described by the vorticity equation. After taking the curl of this equation, we get the vorticity equation for the dust grain medium.
\begin{equation}
\begin{split}
& \frac{d\vec{\omega}}{dt} = \frac{1}{M_d} (\nabla \times \vec{F_g} + \nabla \times \vec{F_E} + \nabla \times \vec{F_i} \\
& + \nabla \times \vec{F_n}) - \frac{1}{M_d n^2_d}\nabla n_d \times \nabla P_E
\end{split}
\end{equation}
where $\vec{\omega} = \nabla \times \vec{v_d}$ is the angular frequency or vorticity vector. Since the gravitational field is curl-free,  the value $\nabla \times \vec{F_g}$ = 0. It is also observed in many experiments that dusty plasma is nearly homogeneous in a 2D plane. In such homogeneous dusty plasma, $\nabla n_d$ $\approx$ 0. With all these approximations, the vorticity equation takes the form
\begin{equation}
\begin{split}
& \frac{d\vec{\omega}}{dt} =
\frac{1}{M_d}(\nabla \times \vec{F_E} + \nabla \times \vec{F_i} \\
& + \nabla \times \vec{F_n})
\end{split}
\end{equation}
It is clear from Eq.9 that vortex generation in a dusty plasma is possible if electrostatic force or ion drag force is nonconservative. In the case of a neutral gas flow, neutral gas friction can also drive a vortex motion. However, if the background gas is at rest, then neutral drag force has only a dissipation effect. It is a fact that coherent or vortex structures in dusty plasma have a finite lifetime; therefore, steady-state solutions of this vorticity equation would describe the onset of such dynamic structures. For the steady state condition, Eq.9 transforms to a given form-
\begin{equation}
\begin{split}
& 0 = (\nabla \times \vec{F_E} + \nabla \times \vec{F_i} \\
& + \nabla \times \vec{F_n})
\end{split}
\end{equation}
%%%%%%%%%%%%%%%%%
\begin{figure} 
\centering
%%%\vspace*{-0.33in}
\includegraphics[scale= 0.70000000]{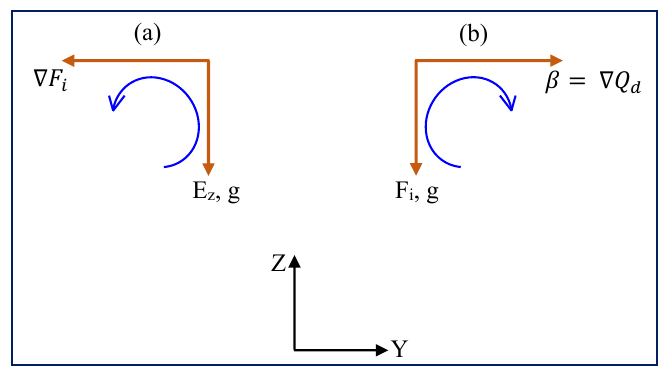}
\caption{\label{fig:fig1} A Schematic representation of the rotational motion in a 2D plane due to (a) ion drag gradient along with orthogonal E-field (b) dust charge gradient ($\beta$) along with the orthogonal non-electrostatic force (gravity). The direction of rotating particles in this Y-Z plane is presented by the blue curve with an arrow.}
\end{figure} 
%%%%%%%%%%%%%%%%%%%%%%%%%%
For a given discharge configuration, we can write electric field force in terms of the plasma potential ($\Phi$) and charge ($Q_d$), $\vec{F_E} = -Q_d \nabla \Phi$. The ion drag force is directed along the E-field and flow of ions. Therefore, the ion drag force vector can be written as $\vec{F_i}$= $F_i \hat{E}$. where $F_i$ is a ion drag force function (magnitude) and $\hat{E}$ represents the direction of ion drag force. The neutral drag force, $\vec{F_n} = -M_d \nu_{dn} \vec{v_d}$.\\\\
Once substitutes the value of forces in Eq.10, we have
\begin{equation}
M_d \nu_{dn} \nabla \times \vec{v_d}   = \nabla Q_d \times \vec{E} + \nabla F_i \times \hat{E} 
\end{equation}
\begin{equation}
\nabla \times \vec{v_d} = \frac{1}{M_d \nu_{dn}}(\nabla Q_d \times \vec{E} + \nabla F_i \times \hat{E}) 
\end{equation} 
There are two possible driving forces or free energy sources $\nabla Q_d \times \vec{E}$ and $\nabla F_i \times \hat{E}$ to sustain the vortex motion of homogeneous and incompressible dust cloud against the frictional losses. In other words, the stable vortex motion of charged micro-particles in the dusty plasma medium can only occur in the presence of free energy sources compensating for the energy losses.\\\\
The first term, $\nabla Q_d \times \vec{E}$ will be non-zero if $\nabla Q_d \neq 0$. This is the case when there is a finite dust charge gradient. Similarly, the second term $\nabla F_i \times \hat{E}$ will be non-zero if $\nabla F_i \neq 0$. It means the gradient in ion drag force, which is not parallel to the E-field, can also excite the vortex motion of the dust cloud. The values of these vortex driving terms strongly depend on the discharge configuration used to produce dusty plasma.\\\\
In most experimental dusty plasmas (finite systems), there are two electric field ($\vec{E}$) components. One is required to hold dust grains against gravity, and the other one helps to compensate for the repulsive forces of dust grains. The ion drag force can have a gradient vertically (along gravity) as well as horizontally (in-plane) or along an arbitrary direction near the power electrode/wall of the chamber. Whenever the value of driving force term $\nabla F_i \times \hat{E}$ is non-zero, then dust grains experience the force that rotates them in the given plane. These rotating dust grains about a common centerline form the stationary vortex structures in that plane. A schematic representation of the rotational motion in the presence of an ion drag gradient orthogonal to the electric field is depicted in Fig.\ref{fig:fig1} (a). \\
In RF capacitively coupled discharge dusty plasma, it is possible to estimate the curl of ion drag force ($\nabla \times \vec{F_i}$) in terms of ion density gradient ($\nabla n_i$) and velocity gradient ($\nabla v_i$) \cite{Chaidustyicevortices2016}
\begin{equation}
  \nabla \times \vec{F_i} = G~\nabla v_i \times \nabla n_i
\end{equation}
where $v_i$ and $n_i$ are the ion drift velocity and ion density respectively. G depends on the grain potential, the radius of the dust particles, ion density, and the ambipolar diffusion coefficient \cite{Chaidustyicevortices2016}. 
The finite curl of ion drag force on dust grains ($\nabla \times \vec{F_i} \neq 0$) is possible if the gradient of $v_i$ is not parallel to the gradient of $n_i$. It should be noted that the formulation (expression) of $\nabla \times \vec{F_i}$ could be slightly different for different dusty plasma systems but the finite value of $\nabla \times \vec{F_i}$ can set the dust grains into rotational motion, resulting in the formation of vortices or coherent structures in the dust cloud. \\\\ 
If we consider a free energy source, $\nabla Q_d \times \vec{E}$ that can excite the vortex motion of dust particles. The spatial dependence of macroparticle charge in a dusty plasma system can convert the potential energy of the electric field into the kinetic energy of dust particles\cite{Vaulinavortices2000,vaulinavortices2004}. The presence of dust charge gradient ($\nabla Q_d$) in the dust-plasma system could be due to nonuniform charging mechanisms because of the inhomogeneity in plasma density and temperature, dispersion of the shape and size of dust grains, etc.\\
Apart from the fluid description, the onset of vortex motion due to spatial dependence of dust charge in the presence of the non-electrostatic force $\vec{F}_{non}$ such as gravitational force ($\vec{F}_g$) or ion drag force ($\vec{F}_i$) can be explained by considering a dust system of finite particles having spatial charge dependence\cite{Vaulinavortices2000,Vaulinavortices2003}. In the presence of a free energy source, the dust particles in a dust cloud start to move in the direction of $F_{non}$ where the dust particle has its maximum charge value and forms a stable vortex structure. In Fig.\ref{fig:fig1}(b), a schematic representation of dust rotation in a 2D plane in the presence of dust charge gradient and orthogonal non-electrostatic forces is displayed.
The frequency ($\omega$) of the steady--state rotation of particles in a vortex structure can be estimated by the formula \cite{Vaulinavortices2000,vaulinavortices2004,Vaulinavortice2010}
 \begin{equation}
 \omega = \vert\frac{F_{non}}{M_d} \frac{\beta}{e Z_0 \nu_{dn}} \vert ,
 \end{equation}
 where $\vec{\beta} = \nabla Q_d = e\nabla Z_d$, $ Z_0 = Q_{d0}/e $ is the charge on the dust particle at an equilibrium position in the rotating plane, $\nu_{dn}$ is the dust-neutral collision frequency.\\
 It should be noted that the discussed theoretical models (\ref{sec:secII} B) help to explore the driving mechanism for vortex formation in the finite dusty plasma systems where the boundary and continuity are dominant ingredients. Therefore, such a theoretical description may not help to understand the driving mechanism of vortex formation in extended dusty plasma systems.\\\\
%%%%%%%%%%%%%%%%
 The vortices in dusty plasma can also arise due to instabilities such as Kelvin-Helmholtz (K-H) instability, Rayleigh-Taylor (RT) instability, dust acoustic wave instability, etc. The Rayleigh-Taylor instability is a kind of buoyancy-driven fluid instability that occurs in the presence of gravitational acceleration ($g$). Whenever a heavy fluid or dusty plasma (high mass density) is supported by a light fluid or dusty plasma in the presence of gravity and a perturbation occurs at the boundary separating both fluids \cite{rtinstability2010,1rtinstabilityavinash2015,rtinstabilityamita2014}, the perturbation grows with time and long-lived mushroom-like vortex structures are formed\cite{rtinstabilityexplaination1,rtinstabilitysanatnature1}.
%%%%%%%%%%%%%%%%%%%%%%%%
\begin{figure} 
\centering
%%%\vspace*{-0.33in}
\includegraphics[scale= 0.400000000]{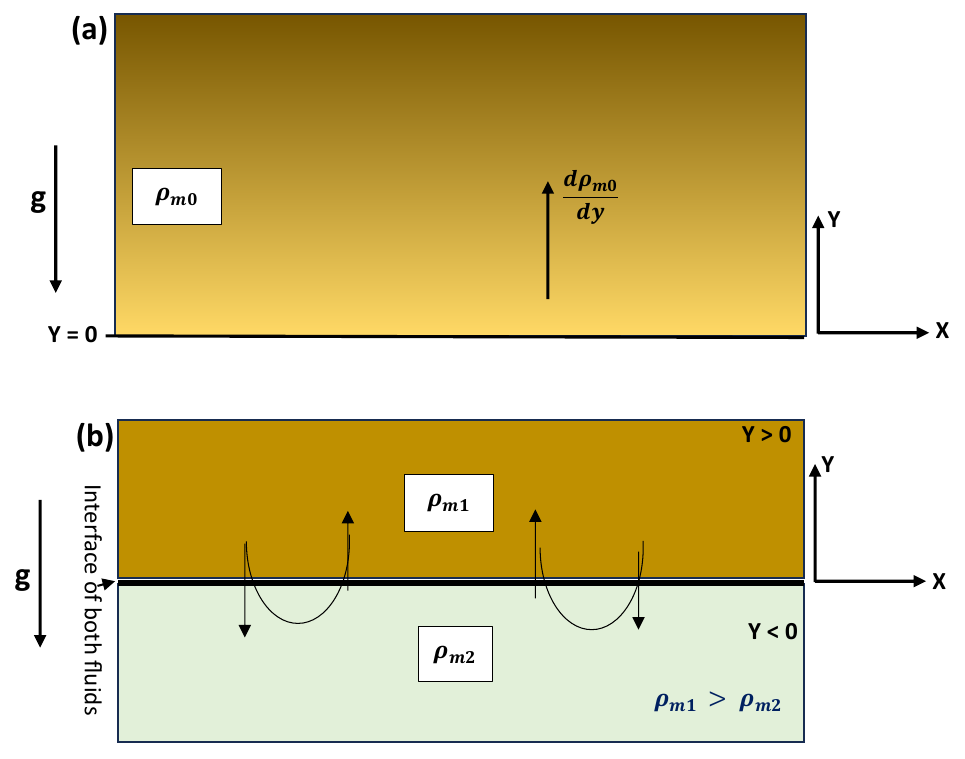}
\caption{\label{fig:fig2} A Schematic diagram of 2D fluid (dusty plasma) (a) having inhomogeneous density variation opposite to the gravitational force, (b) A sharp density variation at the separating boundary of two fluids of different densities/mass densities. The arrows indicate the direction of the flow of fluids during the evolution of RT instability.}
\end{figure} 
%%%%%%%%%%%%%%%%%%%%
A single fluid description of dusty plasma helps in formulating the growth rate of RT instability. The stability of a dust plasma system either in a strongly coupled or weakly coupled state can be understood by the dispersion relation. If strongly coupled inhomogeneous two-dimensional dusty plasma (in X-Y plane), as shown in Fig.\ref{fig:fig2}(a), where dust density increases with height (along y-direction which is opposite to $\vec{g}$) in comparison with the perturbation scale length. The dispersion relation for such a dusty plasma system is \cite{1rtinstabilityavinash2015,rtinstabilityamita2014} 
 \begin{equation}
     \omega^2 = \frac{\eta}{\tau_m} k^2 - \frac{g}{\rho_{mo}} \frac{d\rho_{m0}}{dy} \frac{k^2_x}{k^2}
 \end{equation}
where $k_x$ is a perturbation length scale along the x-axis. $\rho_{m0}$ is the mass density of the fluid (dusty plasma), $\eta$ and $\tau_m$ characterize the nature of dusty plasma. If $\eta \rightarrow 0$ and $\tau_m \rightarrow$ 0 then dusty plasma is considered as simple hydrodynamic fluid (weakly coupled state). For the RT unstable dusty plasma system, the gradient in mass density or dust density gradient should be opposite to the gravity i.e. $\frac{d\rho_{m0}}{dy} > 0$ (as per Fig.\ref{fig:fig2}). The growth rate strongly depends on the strength of the density gradient opposite to the gravitational force acting on the fluid/dust grain medium. The value of the ratio $\frac{\eta} {\tau_m}$ = $\frac{p_d \Gamma}{\rho_{mo}}$ which demonstrates that the growth rate of RT instability reduces with increasing the coupling strength ($\Gamma$) amongst dust grains.\\
If there is a sharp boundary separating two fluids of different densities ($\rho_{m1}$ and $\rho_{m2}$) as shown in Fig.\ref{fig:fig2}(b), then the dispersion relation is \cite{1rtinstabilityavinash2015} 
\begin{equation}
     \omega^2 = \frac{\eta}{\tau_m} k^2 - g k \left(\frac{\rho_{m2}-\rho_{m1}}{\rho_{m1} + \rho_{m2}}\right )
 \end{equation}
In this case, the growth rate of RT instability depends on the wavelength of initial perturbation ($k^{-1}$) as well as the coupling effects. The perturbation grows rapidly for the shorter wavelength. It should be noted that the occurrence of RT instability in either case of dust-plasma system is possible above a critical wave number or initial perturbation wavelength \cite{1rtinstabilityavinash2015}.\\\\
%%%%%%%%%%%%%%%%%%%%%%%%%
\begin{figure} 
\centering
%%%\vspace*{-0.33in}
\includegraphics[scale= 0.70000000]{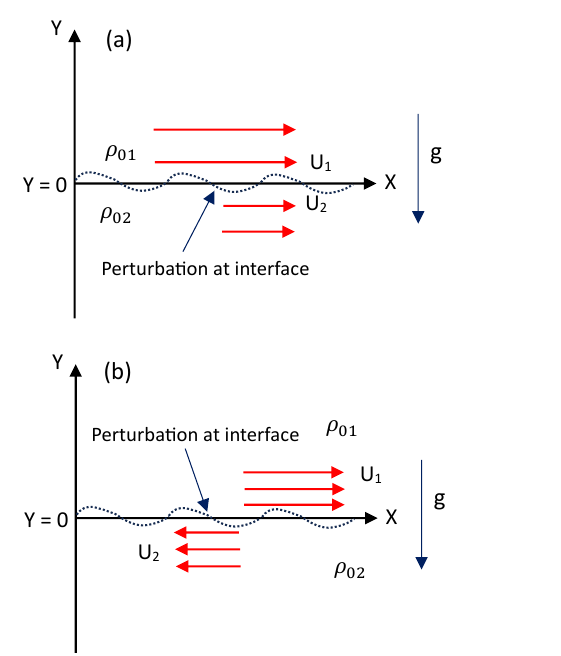}
\caption{\label{fig:fig3} A Schematic diagram of 2-D fluid flow geometry for K-H instability (a) Fluids flowing in the same direction but a finite shear in the velocity at the boundary separating both the flowing fluids (b) Fluids flowing in opposite direction with a finite shear in the velocity at the boundary separating both the flowing fluids. The length and direction of the arrows indicate the magnitude of equilibrium flow velocity and direction of flow respectively. A small linear perturbation at the interface of either flowing fluid is represented by the dotted curve in the figure.}
\end{figure} 
%%%%%%%%%%%%%%%%%%%%%%%
 The K-H instability occurs at the interface of two flowing fluids with different velocities \cite{khinstabilitybanarjeedusty,khinstabilityexplaination1}. Such shear flow can lead to the generation of vortices which is a result of the nonlinear stage of the K-H instability.
 %%%%%%%%%%%%%%%
As per schematic diagram in Fig.\ref{fig:fig3}, two dust grain mediums (fluids) of different mass densities ($\rho_{01}$ and $\rho_{02}$) having a continuous velocity shear in the Y-direction at the boundary of medium (fluid), while flowing either in the same direction (X-direction) or in the opposite directions ($\pm$ X). A set of singly charged fluid equations are solved using the linear stability analysis technique along with some boundary conditions and initial values (incompressible perturbation, variation of perturbed quantities along y-axis)\cite{krall1986principles,khinstability_dolai_2022}. The obtained expression of the dispersion relation is 
\begin{equation}
\omega = k_x (\alpha_1 U_1 + \alpha_2 U_2) \pm \sqrt{-k^2_x \alpha_1 \alpha_2 (U_1 -U_2)^2}
\end{equation}
where $\alpha_1 = \frac{\rho_{01}}{\rho_{01} + \rho_{02}}$ and $\alpha_2 = \frac{\rho_{02}}{\rho_{01} + \rho_{02}}$, $k_x$ is the scale length of perturbation. $U_1$ and $U_2$ are the equilibrium flow velocities of the charged fluid (dusty plasma) in $Y>0$ and $Y<0$ regions respectively as per the schematic representation in Fig.\ref{fig:fig3}\\
In most of the dusty plasma experiments, $\rho_{01}$ = $\rho_{02}$, therefore, the dispersion relation will be
\begin{equation}
\omega = \frac{k_x}{2} (U_1 + U_2) \pm i \frac{k_x}{2} (U_1 -U_2)
\end{equation}
If $\omega$ is real then the linear perturbation at the interface (ref. Fig.\ref{fig:fig3}) will not grow with time and make dusty plasma a system stable. This is the case when no shear in the transverse direction at the separating boundary i.e. $U_1$ = $U_2$. The dusty plasma system is K-H unstable if $\omega$ is an imaginary or complex quantity. This is only possible when there is a finite velocity difference or shear in the velocity ($U_1$ - $U_2$) $\neq$ 0) at the boundary. Once shear velocity ($U_1$ - $U_2$) crosses a critical velocity which can be either the thermal speed or phase velocity of a dusty acoustic wave, instability starts to grow at the interface. Since the vorticity is non-zero ($\nabla \times \vec{v_d}$) due to the discontinuity in velocity at the interface, it induces rotational velocity that can amplify the instability growth. As a result of this instability, vortex rolls or vortices are formed at the interface of the flowing dust grain mediums (hydrodynamic fluids).\\
It should be noted that the growth rate of K-H instability strongly depends on the compressibility in the dusty plasma (compressible dust perturbation). Therefore, the characteristics of vortex structures (shape and size) may get changed while incorporating the compressibility effect of medium\cite{khsanatcompressibility_2012}. When dusty plasma is in the strongly coupled state, its elastic nature also alters the characteristics of coherent or vortical structures \cite{khsanatstronglycopled_2012}.\\\\
%%%%%%%%%%%%%%%%%%%
In dusty plasma, dust-acoustic instability also plays a role in exciting the acoustic wave modes and vortex modes. Such instability arises due to the streaming of plasma ions and neutrals relative to the charged dust particles. The streaming plasma ions/neutrals are assumed to be the free energy sources for the charged dust component against the frictional losses due to the ion-neutral and dust-neutral collisions. The growth of the dust-acoustic instability in the dusty plasma experiments (moderate collisional dusty plasma) is possible above a threshold electric field and increases with the electric field in the dusty plasma. However, the ion-neutral and dust-neutral collisions reduce the growth rate of instability \cite{dustacosutciinstability2,dust-acousticinstability_merlino_1997,dust-acousticinstability_1998,lowerhybridacousticinstability}. Such low-frequency dust-acoustic instabilities \cite{dustacosutciinstability2, lowerhybridacousticinstability} can lead to the formation of vortices\cite{liniacousticvortices2014} as the dust plasma system seeks to attain a more stable equilibrium state.\\\\
%%%%%%%%%%%%%%%%%%%%%%%%%
\begin{figure} 
\centering
%%%\vspace*{-0.33in}
\includegraphics[scale= 0.500000000]{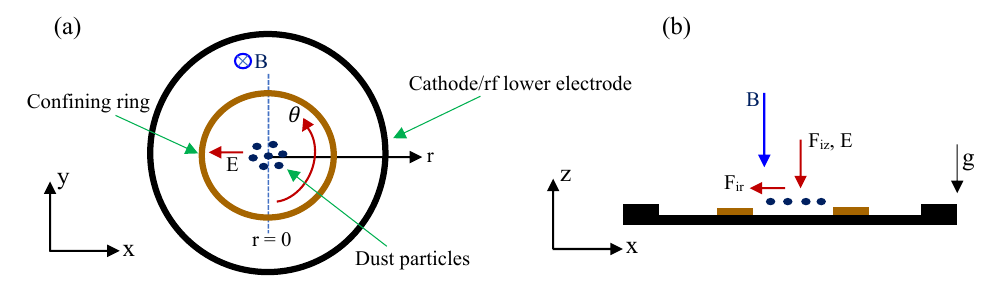}
\caption{\label{fig:fig4} A Schematic diagram of 2D dusty plasma in (a) horizontal plane (X-Y plane) (b) vertical plane (X-Z plane) in the presence of external magnetic field ($\vec{B}$). The arrows indicate the direction of the dust forces/motion in the given plane.}
\end{figure} 
%%%%%%%%%%%%%%%%%%%%%%
In the presence of an external magnetic field ($\vec{B}$), the charged dust particles are set into the rotational motion, forming vortex structures. Similar to the unmagnetized dusty plasma case (B = 0 T), ion drag gradient and dust charge gradient along with E-field (already discussed above) can excite the vortex flow in the presence of external B-field\cite{mangilalvortexbfield2020}.\\
Apart from these driving forces, the $\vec{E}\times\vec{B}$ drift of plasma particles (mostly ions) in the presence of an external B-field set the dust particles into the rotational motion. A schematic diagram in Fig.\ref{fig:fig4} represents the direction of ion drag or E-field in a cylindrical geometry where dust grains are confined by an additional ring against the dust-dust repulsive forces. The ions have radial as well the z-component of velocities in the cylindrical geometry (see Fig.\ref{fig:fig4}(b)). In the presence of B-field, the path of the motion of ions (electrons) is changed from radial to azimuthal direction or in $\vec{E}\times\vec{B}$ direction. This $\vec {E}\times\vec{B}$ drifted ions (electrons) or neutrals transfer the momentum to dust grains and set them into the rotational motion in the plane perpendicular to B-field \cite{Konopkarotation2000,kawrotationwithb2002}. The rotating dust grains about a particular axis form the vortical structures/vortices.\\\\
%%%%%%%%%
\subsection{Boundary conditions and vortices}
\begin{figure} 
\centering
%%%\vspace*{-0.33in}
\includegraphics[scale= 0.440000000]{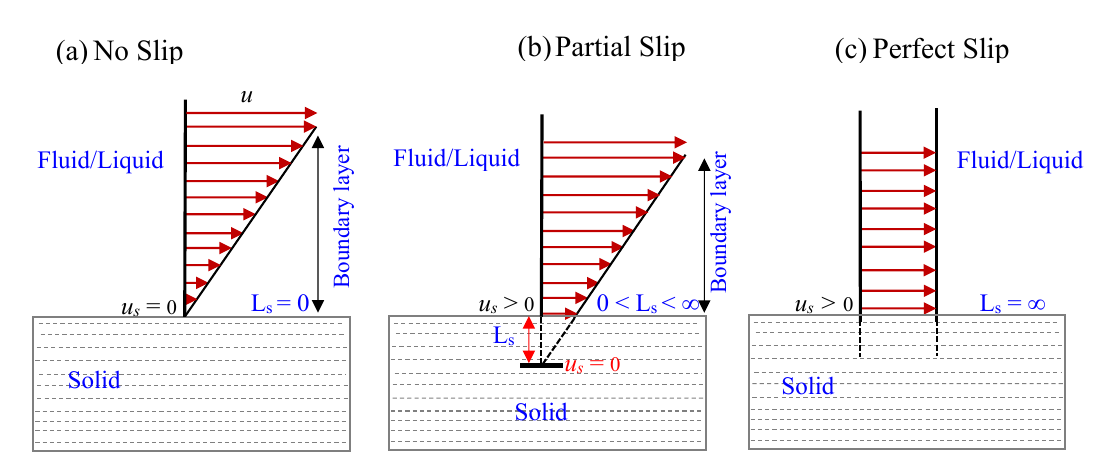}
\caption{\label{fig:fig5} A Schematic diagram representing boundary conditions for hydrodynamic fluid (liquid or gas) flowing over a solid surface. In the diagram, $L_s$ is the slip length, $u_s$ is the velocity at the surface, and $u$ is a constant flow velocity. The boundary layer is very thin or negligible in the perfect slip case.}
\end{figure}

\begin{figure} 
\centering
%%%\vspace*{-0.33in}
\includegraphics[scale= 0.440000000]{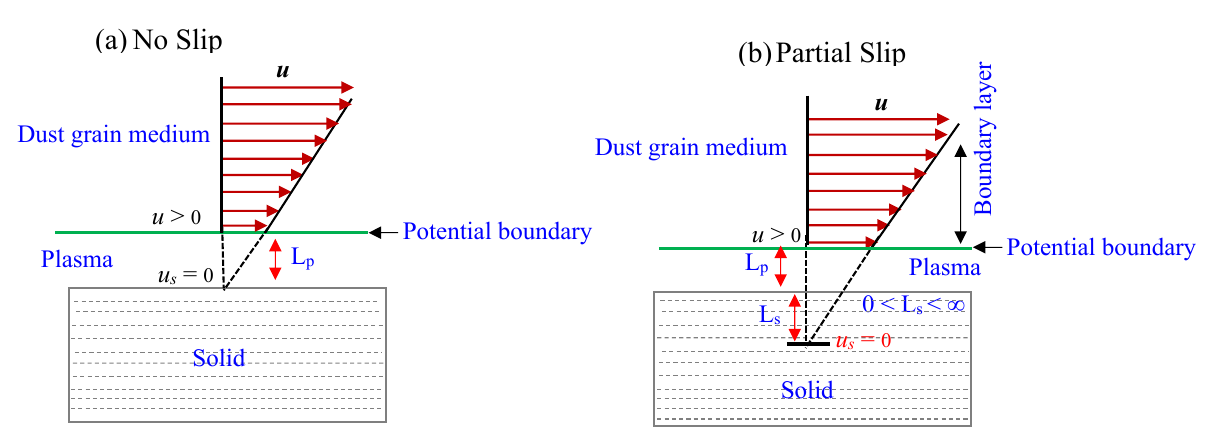}
\caption{\label{fig:fig6} A schematic diagram of dusty plasma flow around the solid surface (a) No slip, (b) partial slip. The potential boundary is separating the plasma and dust grain medium. $L_p$ is the length of the plasma column or sheath dimension or void dimension over the solid surface.}
\end{figure}

It is a fact that the presence (absence) of boundary layers and the nature of boundary conditions (no-slip, partial slip, and perfect slip) play a crucial role in determining the dynamics of vortex formation in the hydrodynamic fluid flows. A schematic representation of boundary conditions (no slip, partial slip, and perfect slip) is depicted in Fig.\ref{fig:fig5}. These boundary conditions affect the distribution of vorticity, the development of shear layers, and ultimately, the formation and evolution of vortices in the flow field\cite{vortexformationobject1,boundaryconditions}. The velocity gradient or shear in flowing fluid around the boundary/surface of the object, which arises due to the non-slip or partial slip boundary (see Fig.\ref{fig:fig5}), determines the strength of shear stress in the tangential direction of flow. The boundary layer thickness (see Fig.\ref{fig:fig5}) depends on various factors such as the medium's viscosity, compressibility, velocity of flowing fluid, mass density of fluid, pressure force, etc. If pressure decreases in the direction of fluid flow, the flow is accelerated. This results in a decrease in the boundary layer's thickness in the flow direction. However, the thickness of the boundary layer increases in the flow direction if pressure increases in the direction of fluid flow. In this case, the boundary layer gets separated from the surface of the body, and counter-rotating vortices are formed in the downstream flow or wake region of the object. The thickness of the boundary layer is predicted by the Reynolds number \cite{vortexformationobject1,raynoldnumbercylinderflow1}, which is a ratio of the inertial forces to viscous forces in the boundary layer. No flow separation occurs at a low value of the Reynolds number, formation of steady vortices or vortex street is possible at an intermediate range of Reynolds number, and flow becomes in turbulent state at a higher value of the Reynolds number\cite{raynoldnumbercylinderflow1,raynoldnumbercylinderflow2,raynoldnumbercylinderflow3}.\\\\
%%%%%%%%%%%%%%%%%%%%%%%%
%%%%%%%%%%%%%%%%%
%%%%%%%%%%%%%%%%%%
In the flowing dusty plasma, the boundary layer or Reynolds number has an effective role on the flow characteristics around the virtual boundary (potential barrier) of an object or void or confining electrode \cite{morfillvoidvortex}. A typical schematic diagram to understand the physical and potential boundary is shown in Fig.\ref{fig:fig6}. The magnitude of dust flow velocity or value of Reynolds number around the virtual boundary (in the boundary layer) of the object/void determines the flow characteristics (laminar or turbulent) and vortex formation in the wake region of the obstacle. The existence of velocity gradient in the tangential direction (shear stress) of dust flow (as per the schematic diagram in Fig.\ref{fig:fig6}) due to the no-slip or partial slip boundary conditions is responsible for the formation of stable vortices and vortex street in the wake region \cite{morfillvoidvortex,harishmdturbulencevortices2016}. However, there is a velocity or Reynolds number range that favors the vortex formation in the wake region of the object \cite{Bailung2020vortex,laserdrivenshearmelzer2012,pintuflowingdustyreview}.\\\\ 
In the bounded dusty plasma, dust grains are confined in an electrostatic penitential well created by the biased and floating electrode/surfaces. The rotating charged dust grains (as per Fig.\ref{fig:fig4}) do not experience any drag/friction force due to the physical boundaries (surfaces/electrodes). However, the potential boundary (virtual boundary) could have a finite role in the flow characteristics of dust grain medium. In such bounded systems, studying the boundary conditions on the vortex motion/flow is challenging because of the potential boundaries rather than physical boundaries (see Fig.\ref{fig:fig6}). The potential boundaries are expected to shift or modify by tuning the plasma parameters. If one changes the potential boundaries, dynamics get altered in response to that dust. This is possible because of the change in dust parameters (charge, forces) associated with the ambient plasma parameters. Therefore, an explicit role of boundaries on the dynamical dust structures in most dusty plasma experiments (bounded system) has not been explored.\\
%%%%%%%%%%%%%%%%%%%%%%%
%%%%%%%%%%%%%%%%%%%%%%%%%%   
%%%%%%%%%%%%%%%%%%%%
\section{Vortices in unmagnetized dusty plasma}   \label{sec:secIII}
In the absence of an external magnetic field, dusty plasma is termed unmagnetized dusty plasma. This section discusses externally driven and instabilities-driven vortices in an unmagnetized dust grain medium. The section is structured based on the driving mechanisms (externally or instability-driven) to form vortices and coherent structures against dissipation losses in unmagnetized dusty plasma.
%%%%%%%%%%%%%%%%%%%%%%
\subsection{External object and force induced vortices}
\begin{figure} 
\centering
%%%\vspace*{-0.33in}
\includegraphics[scale= 0.370000000]{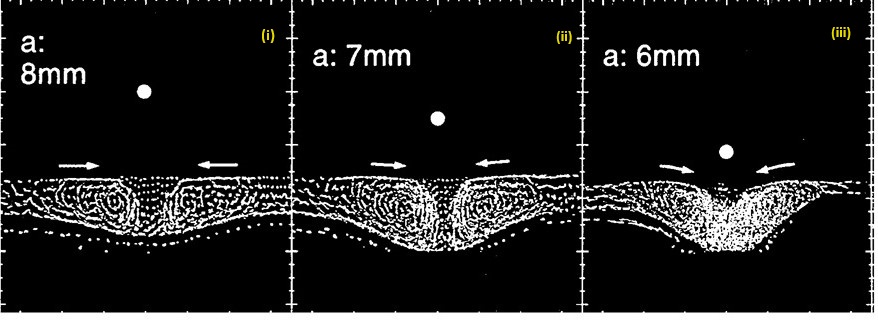}
\caption{\label{fig:fig7}Eighteen overlapping video frames, side views, major ticks 2 mm. During experiments, an RF voltage of 90 V (peak-to-peak) at pressure 0.5 torr was fixed. The probe (wire) was biased at 30 V. Three images (i)–(iii) were taken at different heights (8 mm, 7 mm, and 6 mm) of the probe (white dot in image) from the lower electrode. The biased probe-induced vortex structures are clearly visible in these images. ``Reproduced with permission from Phys. Rev. Lett. 80, 4189–4192 (1998). Copyright 1998 American Physical Society."}
\end{figure}

\begin{figure} 
\centering
%%%\vspace*{-0.33in}
\includegraphics[scale= 0.450000000]{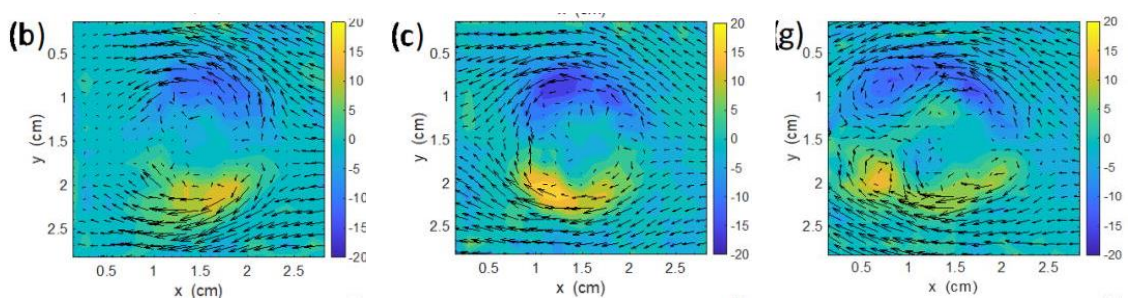}
\caption{\label{fig:fig8}PIV images of dust grain medium at different times (b) (0.63–0.72)s, (c) (0.73–0.82)s, and  (g) (1.13–1.22)s. The dusty plasma was produced in the low power (P = 5W) radio frequency (13.56 MHz) discharge at an argon pressure of $10^{-2}$ mbar. The mono-dispersed silica particles of size 5$\pm$0.1 $\mu m$ were used in these experiments. The color bar in the figure shows the value of vorticity in $S^{-1}$. The vector-less region in images is the location of an obstacle (probe). The dust vortices past the obstacle in the wake region can be identified in the displayed images. ``Reproduced from Physics of Plasmas 27, 123702 (2020), with the permission of AIP Publishing."}
\end{figure}

\begin{figure} 
\centering
%%%\vspace*{-0.33in}
\includegraphics[scale= 0.20000000]{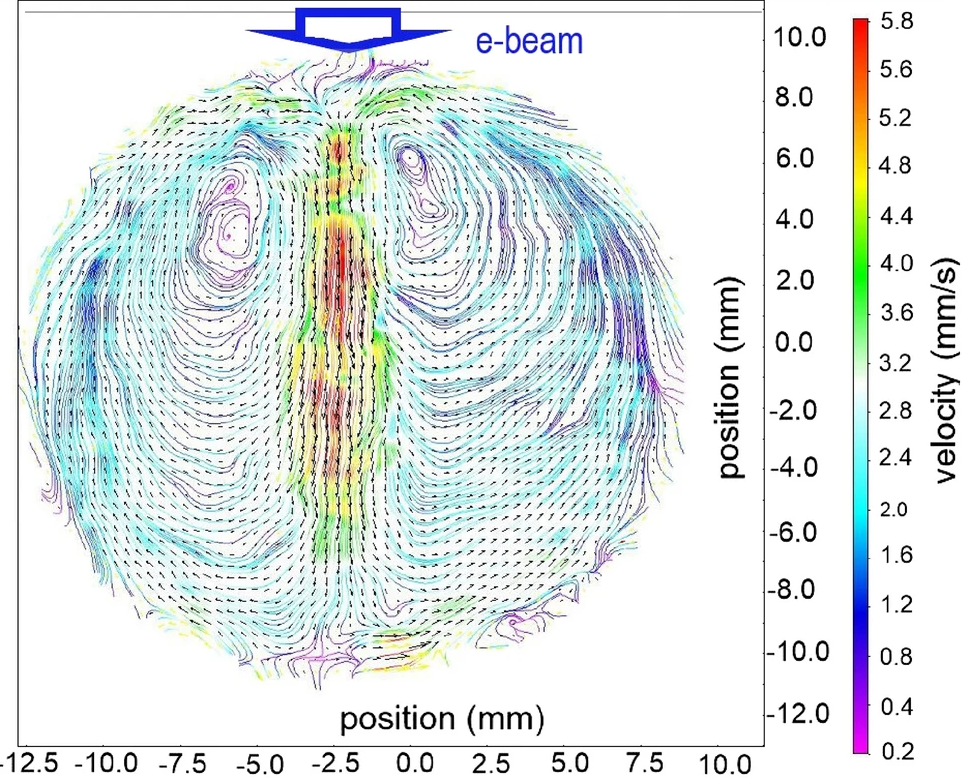}
\caption{\label{fig:fig9}PIV image of the e-beam induced dust flow forming two symmetrical vortices relative to the irradiation direction. The streamlines show the geometry of the flow, while the flow speed is inferred from the color bar. The electron beam of variable acceleration voltage or energy (10 keV - 14 keV) is produced in high vacuum ($10^{-4}$ Torr), while the dusty plasma Crystal is produced at high pressure ($10^{-1}$ Torr) in the capacitively coupled RF (13.56 MHz) discharge. `` D. Ticoş, E. Constantin, M. Mitu, A. Scurtu, and C. Ticoş, Scientific Reports 13, 940 (2023); licensed under a Creative Commons Attribution (CC BY) license."}
\end{figure}

The past study suggests that the vortex motion of charged particles can be excited by introducing external perturbation with the help of a floating or charged object (rod/probe) in dusty plasma. The origin of counter-rotating vortices due to a non-uniform field around a biased probe (metal wire), which is shown in Fig.\ref{fig:fig7}, is reported by Law et al.\cite{probeinduced_vortices_daw}. The dust vortex motion in vertical and horizontal planes was also observed in the presence of an external metal electrode in a planar rf discharge \cite{Samariandustyvorticeselectrode}. The dissipative instability induced by the dust charge gradient and orthogonal non-electrostatic force was considered the primary source of vortex formation. Uchida et al.\cite{uchida2dvortices2009} observed 2-dimensional dust vortex flow in DC discharge near the edge of a negatively biased metal plate introduced externally above the dust levitating electrode. The numerical calculations conclude that asymmetric ion drag force due to the alteration of the potential distribution of confined dust grains in the presence of the biased metal plate plays a dominant role in forming these symmetric dust vortices. 
The experimental study by Jia et al.\cite{singlevortexsawstructure2017} suggests that the vortex motion of dust grains can be set up with the modification in confinement electrostatics potential using an insulator saw structure above the conducting lower rf-powered electrode. The asymmetry of the saw-teeth gives rise to a non-zero curl of the total forces  ($\nabla \times \vec{F} \neq 0$) acting on dust particles which is required for the the vortex formation. In 2020, Dai et al.\cite{vortexmetalsaw2020} reported the vortex formation in unmagnetized dusty plasma by using a metal saw electrode instead of a disk-shaped planar-powered electrode. The asymmetry in azimuthal direction along with the sheath electric field and ion drag was assumed to be a main cause of the vortex motion in such configuration. There is also the possibility of generating rotational vortices in co-generated dusty plasma in the absence of a saw-like electrode \cite{sanjeebdustvortices}. A recent experimental work of Bailung et al.\cite{Bailung2020vortex} suggests that pair counter-rotating vortices can be formed in the wake region behind a stationary obstacle (a metal object) in a flowing dusty plasma medium. The image of vortices past an obstacle in a strongly coupled dust-flowing medium is shown in Fig.\ref{fig:fig8}. They verified the role of the Reynolds number on the formation of vortices behind the obstacle in the wake region.\\ 
%%%%%%%%%%%%%%%%%%%%%%%%
It is possible to modify the dynamical properties of a 2D equilibrium dust grain medium by exerting an external radiation pressure force on dust grains with the help of laser manipulation \cite{laserheatingdustcloud2012,laserdrivenshearmelzer2012}. The dynamical stability of a 2D dust cluster with a finite number of dust particles under the action of external torque was investigated in the laboratory by Klindworth et al.\cite{intershellrotationbylaser2000}. Two opposing laser beams were used to create torque between dust layers with the concept of radiation pressure. They observed that the whole dust cluster rotates as a solid at low laser torque but intershell rotation is possible at higher laser torque. The Monte-Carlo Simulation includes the non-ideal effects to understand the intershell rotation by external lasers torque. A recent experimental investigation demonstrates that an energetic electron beam (8$-$10 KeV) can be used to exert torque on charged particles. The dust particles in a cluster experience torque which is a result of the electron drag \cite{electronbeanrottaionprl2021}. In a large dust cluster (2D dust crystal), a pair of vortices on either side of the electron beam direction are formed with the passing of it through the dust grain medium \cite{Telectronbeamvortices2023}. A PIV image of the electron-beam-induced vortices is depicted in Fig.\ref{fig:fig9}.
%%%%%%%%%%%%%%%%%%%%%%%%%%%
%%%%%%%%%%%%%%
%%%%%%%%%%%%%%%%%%
%%%%%%%%%%%%%%%%
%%%%%%%%%%%%%%%%%%%%%%%%%%%%%
\subsection{Ion drag induced vortex structures}
\begin{figure} 
\centering
%%%\vspace*{-0.33in}
\includegraphics[scale= 0.380000000]{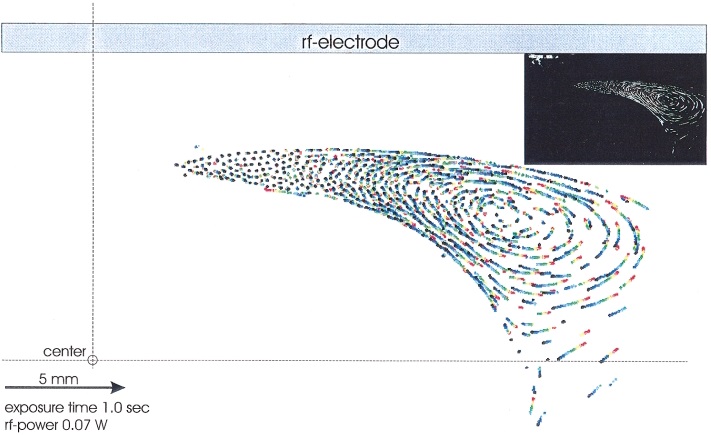}
\caption{\label{fig:fig10} Cross-sectional view through the colloidal plasma condensation in microgravity (TEXUS 35 rocket flight). The rf power was 0.075 W, and 150 video images with a total exposure of 1.0 sec were combined and color-coded, thus tracing the particle trajectories (beginning with “red”). In the inset, the original video image is shown. ``Reproduced with permission from Phys. Rev. Lett. 83, 1598–1601 (1999). Copyright 1999 American Physical Society."}
\end{figure}

\begin{figure} 
\centering
%%%\vspace*{-0.33in}
\includegraphics[scale= 0.57000000]{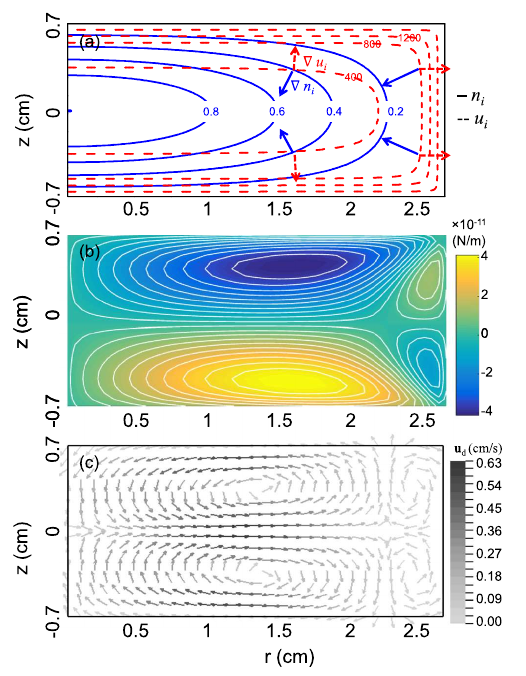}
\caption{\label{fig:fig11}Contour plots of $u_i$ (dashed lines) and $n_i/n_{i0}$ (solid lines). (b)
Contour plot of ($\nabla \times F$). (c) Flow velocity $u_d$ and obtained from numerical
PDE solver. ``Reproduced from Physics of Plasmas 23, 023701 (2016), with the permission of AIP Publishing."}
\end{figure}

It has been experimentally demonstrated that dust grain medium exhibits vortex flow, as depicted in Fig.\ref{fig:fig10}, in RF discharge under micro-gravity conditions \cite{vorticesmicrogravity1999}. The origin of vortex structures observed in microgravity dusty plasma experiments was explained by considering the combined role of ion drag force, electric force,
and screened Coulomb force acting on the dust particles \cite{modelingvorticesmicrogravity1,modelingvorticemicrogravity2,modeingvorticesmicrogravity3}
as discussed in Sec.\ref{sec:secII}B. In a novel ice dusty plasma experiment, Chai et al.\cite{Chaidustyicevortices2016} observed the vortex motion of charged ice particles created in the device by cooling down the water vapor in a controlled manner. The non-conservative nature of ion drag force due to the non-parallel ion density gradient and ion drift velocity gradient (see Fig.\ref{fig:fig11}) acting on charged ice particles set them into vortex motion. 
Manjit et al.\cite{manjeetvortices1,manjeetvortices2} have observed dust vortices (in the 2-dimensional plane) in the poloidal plane of a toroidal geometry in DC discharge configuration. The experimental estimation of forces acting on dust grains in the plasma background confirms the role of the non-conservative nature of the ion drag force in driving vortex motion.    
%%%%%%%%%%%%%%%
%%%%%%%%%%%%%%%%%%%%%%%%%
%%%%%%%%%%%%%%%%%%%
The experimental study of Mulsow et al.\cite{Mulsow20173dvortices} explored the motion of dust grains in a cluster having 50 to 1000 particles using 3-dimensional (3D) diagnostic techniques (stereoscopy). It was observed that dust grains exhibit vortex motion in a poloidal plane when more particles are added to the cluster. The quantitative analysis confirmed the role of the radial gradient of ion drag force in exciting the vortex flow in dust grain medium.
Vladimirov et al.\cite{nucleardustyplasmavortex2001} reported vortex structures with a large number of dust particles in a nuclear-induced dusty plasma system. It was suggested that the motion of ions in the external electric field set dust particles into vortex rotation through an ion-dust momentum transfer mechanism. A ground-based laboratory experiment with the concept of thermophoretic force against gravity on dust particles by Morfill et al.\cite{morfillvoidvortex} confirmed the vortex flow in the wake region of obstacle (here void) in the flowing dusty plasma. To understand the onset and characteristics of vortices around a void in the dust-grain system under micro-gravity conditions, Schwabe et al. \cite{vortexmovementprl20214} conducted a detailed numerical simulation of the 2D dusty plasma system. The addition of extra dust particles to the equilibrium 2D dusty plasma medium makes it unstable, and vortices are formed around the void due to the non-zero curl of the combined ion drag and electric forces acting on particles. A vast analytical and numerical study on the origin of vortex structures in dusty plasma fluid was done by Laishram et al.\cite{Laishramshearflow2014,Laishramstructuretransition2018,Laishramvortices2015,Laishramvortices1,Laishramvortices2,Laishramvortices3} using 2-dimensional (2D) hydrodynamics model. The authors have explored the role of ion shear flow, non-linearity, boundary conditions, and viscosity of the medium in generating a self-organized vortex or coherent structures in the 2D dusty plasma medium. Their study also predicts the formation of multiple co-rotating vortices in the extended 2D dusty plasma medium. In a recent experimental study, Adrian et al.\cite{compressedvortices2023} have observed the stretching and compression of vortices in $CO_2$ rf discharge dusty plasma and discussed the results based on ion drag force as discussed in Sec.\ref{sec:secII}B.
%%%%%%%%%%%%%%%%%%%%%%%%%
%%%%%%
\subsection{Dust charge gradient induced vortices}
\begin{figure} 
\centering
%%%\vspace*{-0.33in}
\includegraphics[scale= 0.20000000]{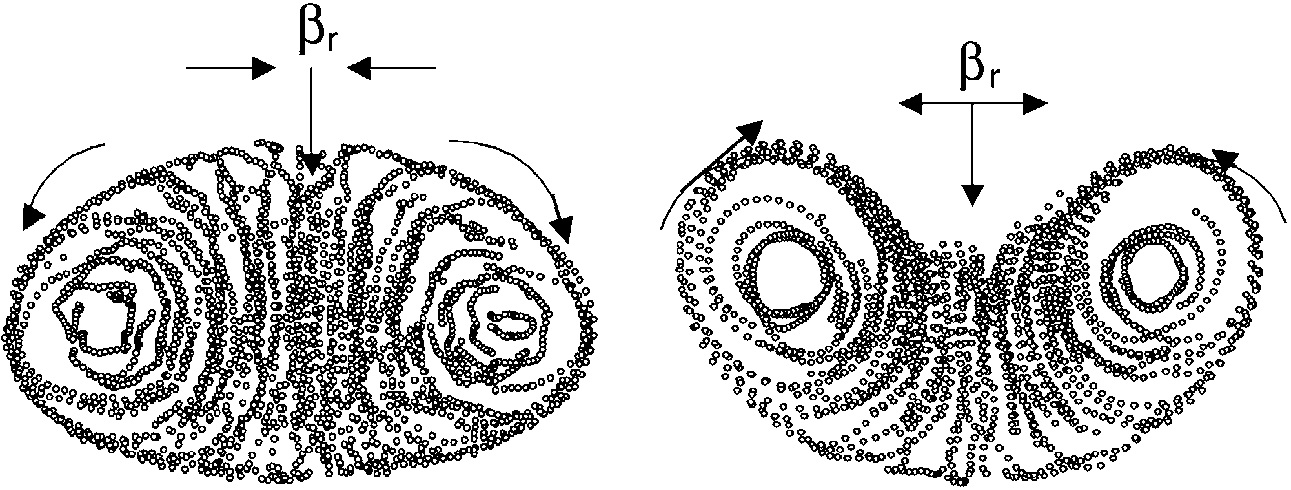}
\caption{\label{fig:fig12} Examples of dust vortex motion (trajectories of particles) obtained from numerical simulations for different parameters (given in original article). $\beta_r$ represents the dust charge gradient direction form the centre of dust cloud and $mg$ is gravitation force acting on particles. The direction of $\beta_r$ decides the direction of vortex flow. ``Reproduced with permission from New Journal of Physics 5, 82.1–82.20 (2003). Copyright 2003 Institute of Physics (IOP). CC BY-NC-SA"}
\end{figure}

\begin{figure} 
\centering
%%%\vspace*{-0.33in}
\includegraphics[scale= 0.3800000000]{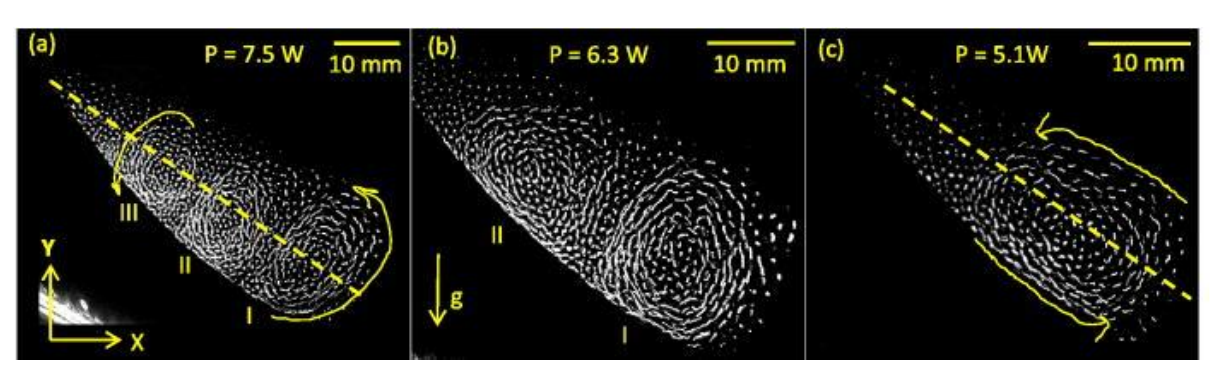}
\caption{\label{fig:fig13}Video images of the dust cloud in the X–Y plane. All images are obtained by the superposition of five consecutive images at a time interval of 66 ms. The vortex structures were observed at different input rf powers. Yellow solid lines with arrow indicate the direction of vortex motion of dust grains and dashed line corresponds to the axis of the dust cloud. ``Reproduced from Physics of Plasmas 24, 033703 (2017], with the permission of AIP Publishing."}
\end{figure}

The self-excited dynamical structures (vortices) in a dusty plasma system having an inhomogeneous plasma background were explored by Vaulina et al. \cite{Vaulinavortices2000,valinaexpvortices2001,Vaulinavortices2003,vaulinavortices2004,Vaulinavortice2010}. The role of dust charge gradient along with orthogonal non-electrostatic force such as gravitational or ion drag force in such dissipative dusty plasma medium in the formation of vortices were discussed in Sec.\ref{sec:secII}B. A typical dust vortex flow pattern in the presence of dust charge gradient ($\beta_r$) along with gravitational force ($mg$) is presented in Fig.\ref{fig:fig12}. The mathematical model provided in Sec.\ref{sec:secII}B had helped to explain the experimentally observed dust vortices in planar rf discharge dusty plasma\cite{Samariandustyvorticeselectrode,valinaexpvortices2001}. In 2003, Agarwal et al.\cite{agarwalrotation} observed the rotating structures (vortices) in DC discharge strongly coupled dusty plasma without any external metal electrode. The vortex flow in such a dissipative dusty medium was understood with the help of a theoretical model provided in Sec.\ref{sec:secII}B.  Ratynskaia et al. \cite{Ratynskaialargescalevortices2006} have performed the experimental study of dust grains in a 2D mono-layer complex plasma and reported a wide range of vortical flows or structures. One of the causes for the formation of large-scale vortices in 2D complex plasma was assumed to be charge inhomogeneity across the dust layer. A recent experimental study by Choudhary et al.\cite{mangicorotating2017,mangivortex2018} has confirmed the dust charge gradient along with ion drag force or gravity as a driving force to excite the vortex motion against dissipative losses in a dusty plasma medium. In this work, a novel experimental configuration using inductively coupled discharge was proposed to create a large aspect ratio dusty plasma with an inhomogeneous plasma background that can accommodate multiple self-sustained vortex structures \cite{mangicorotating2017,mangivortex2018}, as is displayed in Fig.\ref{fig:fig13}.
%%%%%%%%%%
%%%%%%%%%%%%%%%%%%%%%%%%
%%%%%%%%%%%%
%%%%%%%%%%%%%%%%%
%%%%%%%%%%%%%%
%%%%%%%%%%%%%
\subsection{Neutral drag driven vortex structures}
\begin{figure} 
\centering
%%%\vspace*{-0.33in}
\includegraphics[scale= 0.6000000]{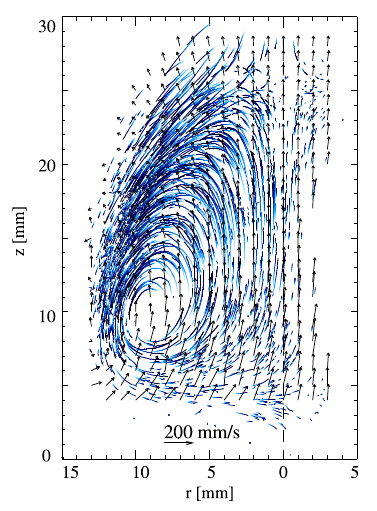}
\caption{\label{fig:fig14} Averaged gas flow velocity field (vectors) superimposed with particle trajectories. The vertical dashed line indicates the center of the tube. ``Reproduced with permission from
Phys. Rev. Lett. 101, 235001 (2008). Copyright 2008 American Physical Society."}
\end{figure}

In 2008, Mitic et al.\cite{miticvortices2008} first time observed vortex structures (see Fig.\ref{fig:fig14}) in a dust grain medium confined in a vertical glass tube due to the thermal creep of background atoms/molecules. In their study, the gas flow driven by thermal creep was anti-parallel to gravity. The background gas atoms driven by thermal creep exert a drag force on dust grains, forming vortices in dust grain medium \cite{Schwabethermalvortices2010}. The experimental study by Flanagan and Goree \cite{thermalcrepdustvorticeflagna2009} also demonstrated the role of thermal creep flow in setting up the dust grains into rotational motion. The temperature gradient along a surface in contact with a low-pressure gas creates thermal creep flow, resulting in the formation of 2D vortex flow due to the easy response of dust particles with ambient gas flow. A recent experimental study under micro-gravity conditions \cite{thremalcreepvortices2023} also confirms the role of thermal creep in driving the dust grains into rotational motion, resulting in counter-rotating vortices. It has been reported \cite{magnetizationwithneutralgas2012} that frictional coupling between dust particles and background neutral gas can be used to realize the magnetization of massive charged dust particles. The dust grains behave in a way that is equivalent to being magnetized in the absence of an external magnetic field. The directed flow (rotational flow) of background gas sets dust grains into rotational motion through momentum transfer mechanisms \cite{magnetizationwithneutralgas2012}. If sufficient dust grains are used in the experiment, dust vortices are expected to form in a strongly coupled dusty plasma medium by rotating neutral gas. The role of frictional coupling of dust particles with background neutrals in exciting the large-scale shear modes such as vortices or inter-shell rotations in the Yukawa balls (dusty plasma ball) was explored experimentally by Ivanov and Melzer\cite{yukawaballvorticesexp2009}. They used singular value decomposition and normal mode analysis to explore the dynamical features of Yukawa balls. Their analysis predicted various modes but large-scale vortices or intershell rotations were dominant modes. In computer experiments, it is possible to create a temperature difference between charged dust layers and study the effect of temperature gradient on dust dynamics. With this objective, Charan et al.\cite{harish2015} performed a molecular dynamics simulation of strongly coupled dusty plasma with a temperature difference between charged dust layers and observed the vortex flow.
%%%%%%%%%%%%%%
%%%%%%%%%%%%%
\subsection{Instabilities induced vortices}
\begin{figure} 
\centering
%%%\vspace*{-0.33in}
\includegraphics[scale = 0.5500000]{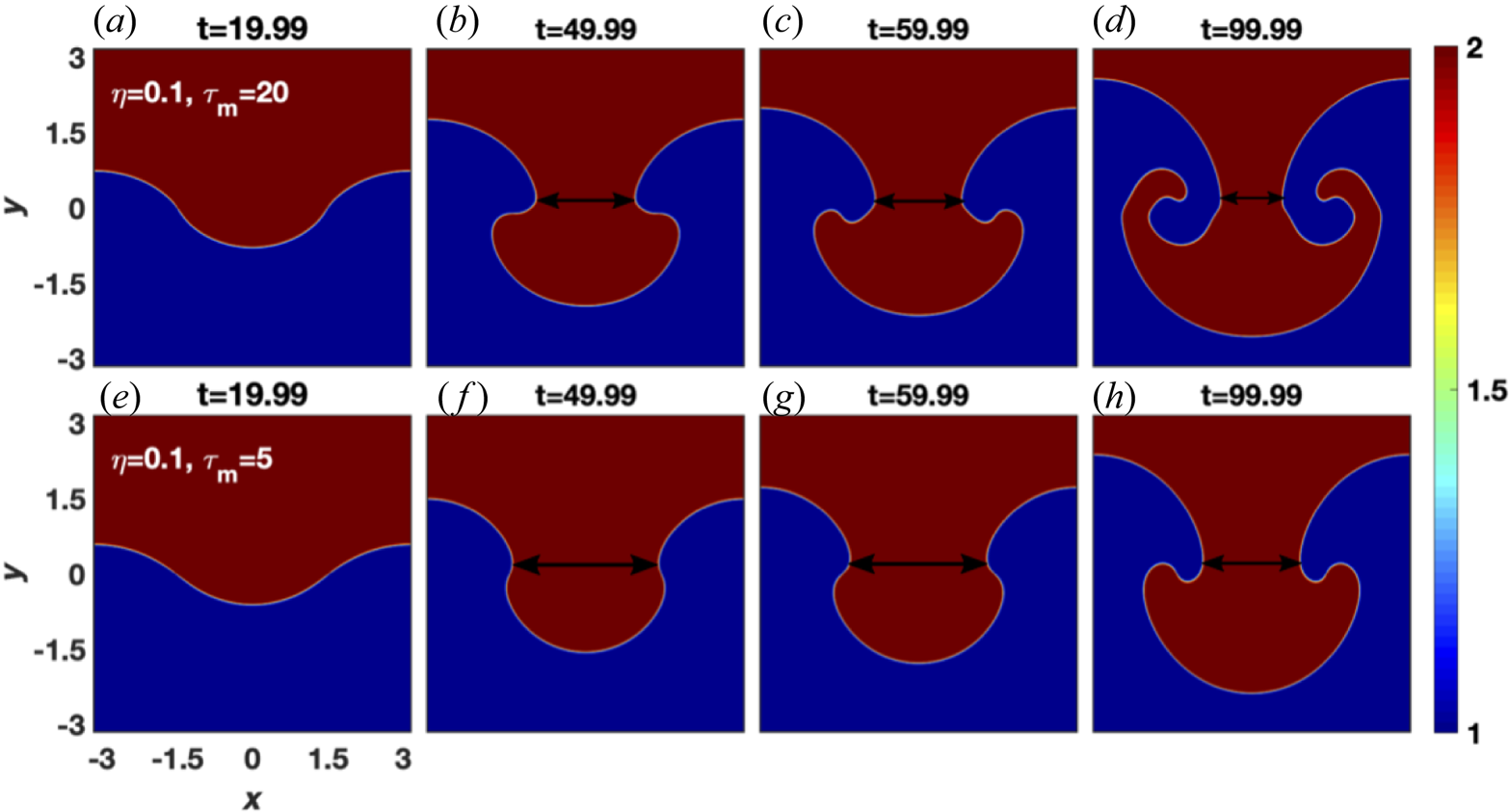}
\caption{\label{fig:fig15} The growth of RT instability at the sharp interface of two visco-elastic fluids (dusty plasma) of different densities: (a–d) $\eta$ = 0.1,  $\tau_m$ = 20; $\eta$ = 0.1,  $\tau_m$ = 5. ``Reproduced with permission from Journal of Plasma Physics 87, 905870216 (2021). Copyright 2021 Cambridge University Press."}
\end{figure}

The role of various instabilities in exciting the vortices in the finite or extended dusty plasma systems has been discussed in Sec.\ref{sec:secII}B. Veeresha et al.\cite{rtdrivenvortices2005} demonstrated the Rayleigh–Taylor instability (RT-instability) driven flow patterns or vortices in dusty plasma. These vortex structures or modes were observed in the saturation state of the RT instability. The numerical work of Chakrabarti et al.\cite{rtvortices2} also confirmed that the nonlinear evolution of RT instabilities can form coherent vortex structures in dusty plasma. The stability of such coherent vortex structures depends on the strength of short-range secondary instability, which is excited in the core of vortex \cite{rtvortexstability}. Recent numerical simulations by Dharodi et al.\cite{vikramrtvortices} also observed the vortex-like structure as shown in Fig.\ref{fig:fig15} during the time evolution of RT instability in visco-elastic dusty plasma fluid. However, no experimental work has been reported to confirm the RT instabilities that drive coherent vortex structures in dusty plasma.\\
%%%%%%%%%%
%%%%%%%%%%
\begin{figure} 
\centering
%%%\vspace*{-0.33in}
\includegraphics[scale= 0.3000000]{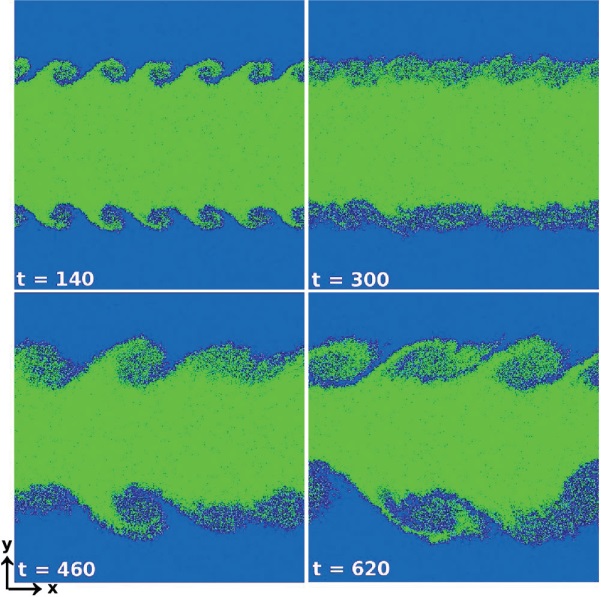}
\caption{\label{fig:fig16} Blue colored fluid moves in the $+$ x and green colored moves in $-$x. Inverse cascading of mode $m_n$ = 6 starting from an initial state of coupling constant 50. Formation of vortices and their evolution due to K-H instability can be seen in images. ``Reproduced with permission from
 Phys. Rev. Lett. 104, 215003 (2010). Copyright 2010 American Physical Society".}
\end{figure} 
%%%%%%%%%%%%%%
Another kind of instability that has been studied for a long time in dusty plasma is Kelvin Helmholtz (K-H) instability\cite{khinstabilitybanarjeedusty}. Ashwin et al.\cite{ashwinkhvortices2010} performed molecular dynamic (MD) simulation to explore the K-H instability in strongly coupled dusty plasma. Time evolution of K-H instability produces vortex roll in the nonlinear regime depicted in Fig.\ref{fig:fig16}. Numerical and analytical studies of K-H instabilities driven flow patterns (vortex structures) in its nonlinear regime in weakly as well as strongly coupled dusty plasma has been studied by various research groups \cite{sanatkhvortices2012,sanatkhvortices2,dharodikhvortices2022,sanatkhvortices3,das_dharodi_tiwari_2014}. The analytical study by Janaki et al.\cite{janakishearedinducedvortex2010} had claimed the existence of dipolar vortex structures in 2D dusty plasma medium in the presence of sheared dust grains flow. Gupta et al.\cite{Guptavortices1,Guptavortices2} performed the computational fluid dynamics and molecular dynamics simulations of dusty plasma medium with Kolmogorov shear flows and observed the coherent vortex structures along with other flow patterns as a result of the sheared flow-induced instabilities. A recent experimental study by Krishan et al.\cite{kishankhvortices2023} confirmed the K-H instability-driven vortex structures in DC discharge flowing dusty plasma. They generated shear at the interface of the moving and stationary layers of dust grain medium that excites the K-H instability, and evolution of it gives vortex structures, as is shown in Fig.\ref{fig:fig17}. 
%%%%%%%%%%%%%%%%%
\begin{figure} 
\centering
%%%\vspace*{-0.33in}
\includegraphics[scale= 0.3000000]{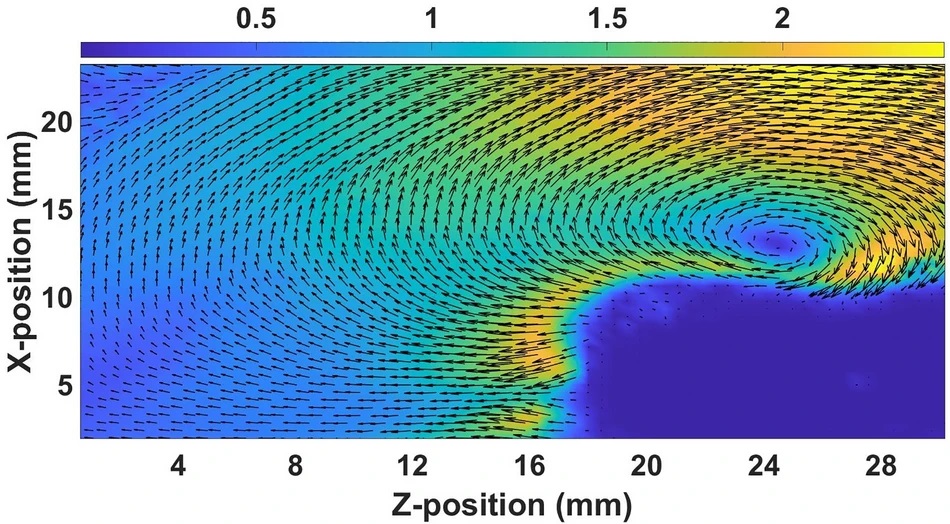}
\caption{\label{fig:fig17}The velocity vector field with the magnitude of the velocity ($cm/s$) for the double layer flow at the input pressure of the pulse valve is 300 Pa. The vortex pattern in the displayed image arises due to the K-H instability. ``K. Kumar, P. Bandyopadhyay, S. Singh, V. S. Dharodi, and A. Sen, Scientific Reports 13, 3979 (2023); licensed under a Creative Commons Attribution (CC BY) license."}
\end{figure} 
\section{Vortices in magnetized dusty plasma} \label{sec:secIV}
%%%%%
The previous section deals with vortex and coherent structures in unmagnetized dusty plasma where different free energy sources to drive vortex motion against dissipation losses were elaborated in detail. Since dusty plasma is an admixture of electrons, ions, and negatively charged dust grains, an external magnetic field strongly affects their dynamics through magnetic or Lorentz force [$F_m = q (\vec{v} \times \vec{B}$)]. The strength of Lorentz force strongly depends on the velocity (or energy and mass) of the charged particle; therefore, lighter charged particles (electron/ions) having higher velocity experience a strong magnetic force as compared to the massive slow-moving charged particles (dust grains) at a given strength of the magnetic field. Hence, the dynamics of electrons and ions get modified at a weak B-field, but a strong B-field is required to make Lorentz force effective for charged dust particles \cite{thomasmagnetized2015,melzermagnetizedreview2021}. There are many challenges to achieving magnetization conditions for charged dust grains\cite{thomasmagnetized2015,mangicppreview2023} in the laboratory, therefore, most past studies on magnetized dusty plasma, where only ambient plasma species (electrons and ions) are magnetized were conducted. The role of the external magnetic field in establishing the vortex structures through the dynamics of plasma species in the dust plasma system is discussed in this section.     
%%%%%%%%
\subsection{$\vec{E}\times\vec{B}$ drift induced vortex motion}
%%%%%%%%%
\begin{figure} 
\centering
%%%\vspace*{-0.33in}
\includegraphics[scale= 0.5000000]{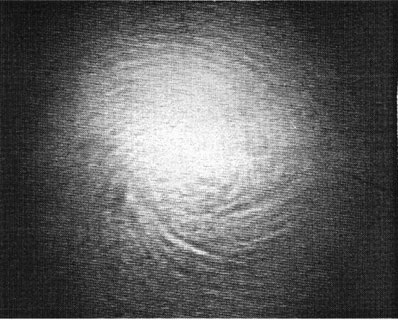}
\caption{\label{fig:fig18}Experimental spiral vortex pattern (Power = 40 - 50 W, p = 120 Pa). ``Reproduced with permission from Plasma Science and Technology 9, 11–14 (2007). Copyright 2007 IOP Sciences."}
\end{figure} 
%%%%%%%%%%%%%
In the presence of an external B-field, $\vec{E}\times\vec{B}$ drifted ions (electrons) exert force on dust grains and set them into rotational motion \cite{Satodustrotationinmagneticfield,Konopkarotation2000,asanrotationinbfirled2019,kawrotationwithb2002,Dzlievadustrotation2016} or background neutral atoms is set into rotation by azimuthal ion flow, and then dust particles start to rotate by this neutral gas flow \cite{gasflowdrivenvortex,shukladustvortexmodel2002}. The trajectories of the rotating dust grains can be rigid rotation \cite{dustytorusanodepiel2010,dusttorusinbrigidrot2015} or shear rotation \cite{mangiannulardusty,Konopkarotation2000} or vortex rotation \cite{mangilalvortexbfield2020,Vasilievdustvortexwithb2011}. Huang et al.\cite{vortexwithbhaung2007} observed the various vortex patterns of dust grains in magnetized dusty plasma experiments (See Fig.\ref{fig:fig18}) where the external B-field was 0.2 T. They observed the gathering and dispersive vortex structures in a 2D plane after tuning the input RF power. The molecular dynamic simulation and analysis of forces acting on dust grains confirmed the role of confining electric potential along with strong B-field in the transition of circular rotating dust particles to vortex strictures. To understand dust vortex patterns, Nebbat et al.\cite{Nebbatvortexwithb2010model,Nebbat2023} did calculations using the time-dependent nonlinear model for such dusty plasma system and highlighted the role of magnetized ions having azimuthal velocity component in driving vortex motion. Vasiliev et al.\cite{Vasilievdustvortexwithb2011} presented experimental results on rotating dust structure (vortex) in a horizontal plane of a cylindrical device with DC discharge configuration under action of strong B-field up to 0.25 T. The rotation of dust grains in the structure was explained by calculating the ion drag force acting on particles against friction force arising due to the background neutrals. In a strong B-field (B $>$ 0.5 T), dust clusters in the 2D horizontal plane also exhibit vortex-like motion that was confirmed in MDPE dusty plasma device \cite{surabhimagnetized2017}. In 2013, Saitou et al.\cite{saitodustdynamobfield2013} studied the dynamics of dust grain medium in the presence of a strong B-field. They observed the transition from horizontal rotational motion to vortex motion in the vertical plane of a cylindrical system once the strength of the external B-field is increased. The disturbance of the magneto-hydrodynamic equilibrium of dust grain medium at a higher value of B-field was considered the leading cause of the vortex motion of dust grains in the vertical plane. It should be noted that dust clouds that rotate as a whole at low to moderate magnetic fields in a horizontal plane break up into smaller vortices at high magnetic fields (few Tesla). Schwabe et al.\cite{magneticfilamentationdusty2011} performed the experiments in low-pressure dusty plasma and observed small-scale rotating or vortex structures in a 2D dusty plasma at strong B-field (B $>$ 1 T). The breaking up of rotating 2D dust medium into multiple rotating structures is a result of the plasma filamentation at a strong magnetic field\cite{magneticfilamentationdusty2011, Thomas_2020}.  
%%%%%%%
\subsection{Sheared flow and dust charge gradient driven vortices}
\begin{figure} 
\centering
%%%\vspace*{-0.33in}
\includegraphics[scale= 0.6000000]{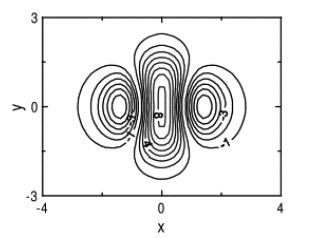}
\caption{\label{fig:fig19}The tripolar vortex for given perturbed potential and other parameters [follow the original paper\cite{tripolarvorticesmagnetizeddusty2001}]. Two lateral vortices have opposite direction
of rotation with respect to the central vortex. ``Reproduced with permission from Physics Letters A 278, 231–238 (2001). Copyright 2001 Elsevier."}
\end{figure}

The analytical study using a fluids model conducted by Bharuthram et al.\cite{shuklavorticsmagnetized1992} of non-uniform dusty plasma in the presence of B-field predicated dipolar vortices as a result of the stationary solution of nonlinear equations. Their study demonstrates that shear in ion flow excites low-frequency electrostatic modes, which interact with and give rise to vortex structures in dust plasma systems. In an external magnetic field, the equilibrium sheared plasma flow or nonuniform plasma is considered a free energy source (as discussed in Sec.\ref{sec:secII}B) to excite various linear and nonlinear modes in the medium. The saturation state of nonlinear coupling of excited modes gives various vortex structures such as vortex chain, tripolar vortices (see Fig.\ref{fig:fig19}), and global vortices in magnetized dusty plasma system \cite{tripolarvorticesmagnetizeddusty2001}. It is also possible to excite dipolar vortices in magnetized dusty plasma because of the nonlinear saturation of K-H instability \cite{khinstabilitymagneticdusty1993,shukladustyvorticesmagnetizeddusty1993,dipolarvortices2}. The role of the sheared magnetic field that can produce sheared plasma flow perpendicular to the B-field on dusty plasma was examined by Jovanovic et al.\cite{diploarandtripolarvortices2001}. The authors observed the formation of dipolar and tripolar vortices depending on the initial plasma density profiles in non-uniform magnetized dusty plasma.\\
%%%%%%%%%%%%%%%%
%%%%%%%%%%%%%%%%%%
\begin{figure} 
\centering
%%%\vspace*{-0.33in}
\includegraphics[scale= 0.5000000]{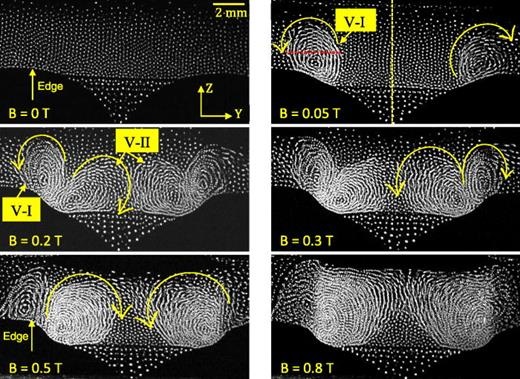}
\caption{\label{fig:fig20}Video images of the dust cloud (aluminum ring) in the vertical (Y–Z) plane. Images at different magnetic fields are obtained by a superposition of five consecutive images at a time interval of 65 ms. The edge vortex and central region vortex are represented by V–I and V–II, respectively. The vortex structures at different strengths of the magnetic field are observed at a fixed input rf power, P = 3.5 W, and argon pressure, p = 35 Pa. The dotted yellow line represents the axis of symmetry. The yellow solid line with an arrow indicates the direction of the vortex flow in the vertical plane of the 3D dusty plasma. ``Reproduced from Physics of Plasmas 27, 063701 (2020)], with the permission of AIP Publishing."} 
\end{figure} 
%%%%%%%%%%%%%%%%%
Dasgupta et al.\cite{dasguptavorticesmagnetized2021} investigated the dust dynamics numerically in the presence of external B-field and predicted vortices in dust plasma system by considering the role of dust-dust interactions, dust diffusion, and ion drag force in the formation of vortex structures. A recent numerical simulation by Prince et al.\cite{princevortexmagnetized2020} examined the dynamics of a dusty plasma medium in a weakly non-uniform magnetic field. The non-uniformity in the external B-field induces $\vec{E} \times \vec{B}$ drift flow of plasma particles. The sheared $\vec{E} \times \vec{B}$ drift ions play a major role in driving the vortex motion as discussed in Sec.\ref{sec:secII}B in a magnetized dusty plasma system. Laishram \cite{laishramvorticesmagnetized2021} further extended the study of dusty plasma in non-uniform external B-field and observed the sheared between electrons and ions due to $\vec{E} \times \vec{B}$ and $\nabla B \times \vec{B}$ drifts in the presence of B-field. The sheared nature of these drifts was considered the vorticity source to establish the vortex structures in dusty plasma. It was also numerically investigated that temperature gradient in background ions plays a dominant role in forming vortices in magnetized dusty plasma \cite{haque_saleem_2006}. In recent experimental work, Choudhary et al.\cite{mangilalvortexbfield2020} reported the existence of a pair of counter-rotating vortices, as shown in Fig.\ref{fig:fig20}, in the vertical plane of a 3D dusty plasma in the presence of strong B-field (B $>$ 0.05 T). The vortex structures originating in dust grain medium were explained by the combined effect of dust charge gradient and iron drag force gradient along with gravitational force/electric force in the cylindrical system. An analytical work by Shukla and Mamun et al.\cite{shukla_2003_vortices_magnetoplasma} reported the electrostatic and electromagnetic vortices in a non-uniform magnetoplasma containing immobile dust grains. The evolution of nonlinear dispersive solitary modes (electrostatic and electromagnetic) explains the origin of dipolar vortices in such dusty plasma systems.       
%%%%%%%%%%
\section{Evolution of Vortex patterns} \label{sec:secV}
\begin{figure} 
\centering
%%%\vspace*{-0.33in}
\includegraphics[scale= 0.7000000]{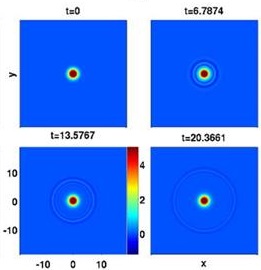}
\caption{\label{fig:fig21} Evolution of smooth circular vorticity profile in time for visco-elastic fluid (dusty plasma). ``Reproduced from Physics of Plasmas 21, 073705 (2014)], with the permission of AIP Publishing."}
\end{figure}

%%%%%%%%%%%
The vortex development and evolution in a normal fluid are entirely different from that observed in a dusty plasma medium considered a visco-elastic fluid. The stability of vortices or coherent structures in the inviscid fluid is perfectly stable over time but the destabilization of vortices is expected in a dusty plasma medium. It should be noted that the concept of inviscid flow is a simplification, of course, in the case of conventional fluids. Several factors can affect the stability of vortices. For example, the viscosity of the medium, interaction strength between charged grains, external perturbation, changes in the flow regime (laminar to turbulence flow), the interaction of vortices, and instabilities develop in the system. In recent years, numerous authors have performed analytical and computational works to explore the evolution, interaction, and stability of dust vortices. Dharodi et al.\cite{vikramvortice1} performed numerical simulation to explore the evolution and interaction of coherent vortices in strongly coupled dusty plasma by considering its visco-elastic nature. They observed the originating transverse shear waves during the evolution of vortex patterns as is depicted in Fig.\ref{fig:fig21}. 
%%%%%%%%%%%%%%%%%
\begin{figure} 
\centering
%%%\vspace*{-0.33in}
\includegraphics[scale= 0.60000000]{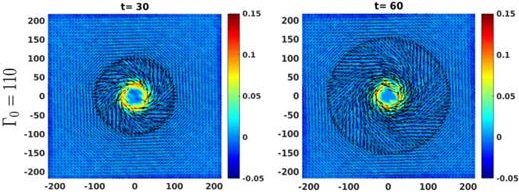}
\caption{\label{fig:fig22} Contour plot of fluid vorticity obtained from MD simulation. Black colored arrows show the velocity field. The grain velocities in the bins are fluidized through a 55$\times$55 grid to construct vorticity. ``Reproduced from Physics of Plasmas 27, 050701 (2020)], with the permission of AIP Publishing."}
\end{figure} 
%%%%%%%%%%%%%%%%%%
The role of collision on the interaction and evolution of counter-rotating and co-rotating vortices in such dusty medium was examined by numerical simulation \cite{Janavorticestheory}. The effect of viscoelasticity and compressibility on the vortex patterns and originating inertial waves due to vortex motion (see Fig.\ref{fig:fig22}) were explored using computer experiments by Gupta et al.\cite{akankshavorticeevolution12019,akankshavorticesevo2020,Guptareview2022} in strongly coupled state of dust grain medium. The interaction of coherent vortex structures at the interface of two inhomogeneous dusty plasma mediums gives birth to transverse shear as well as spiral density waves \cite{vikramvortices2}. The large-scale MD simulation of a strongly coupled dust plasma system predicted the emergence of isolated coherent tripolar vortices during the evolution of unstable axisymmetric flows \cite{ashwintripolarvortices_2011}. A classical MD simulation study to explore the formation and collisions of dipolar vortices in 2D strongly coupled dusty liquid was carried out by Ashwin and Ganesh \cite{ashwindiploarvorticescollison2012}. In their setup, the dipolar vortices emerged from a sub-sonic jet profile in a less dissipative dust grain medium. During a centered head-on collision between dipolar vortices, the vortex patterns are exchanged, not mass. It is a fact that circular vortices are stable due to energy balance. However, elliptical vortices can be formed in a dusty plasma medium if circular vortices deviate from their shape. The deformed dust vortices take an elliptical shape, which is unstable. The stability of elliptical vortices and the formation of new small-scale vortices in dusty plasma were studied by Jana et al.\cite{ellipticalvortices2015}. It should be noted that this section does not discuss the evolution and stability of vortices, which were mentioned in Sec.\ref{sec:secIII} and Sec.\ref{sec:secIV}. This section highlights some independent research works with exciting results (as discussed) during the vortex evolution. 
%%%%%%%%%%%%%%%%%
\section{Turbulence and vortices} \label{sec:secVI}
%%%%%%%%%%%%%%%
In the fluids, turbulence is accompanied by vortices or vortex flow under certain flow conditions. Small-scale vortices can also be formed in a turbulent flow of plasma or dusty plasma medium. The study of the development of turbulence in the dusty plasma at low Reynolds number was performed by Schwabe et al.\cite{turbulenceschawabe2027}. The authors discussed the role of instability (heart-beat instability) and auto-oscillation in developing turbulence in the dust-grain system. To understand the turbulence in the wake region of an obstacle at the kinetic level, Joshi et al.\cite{joshiturbulencevortices2024} performed a 3D molecular simulation of complex plasma flowing past an obstacle. The obstacle was not a physical object but stationary negative charges that could repel the dust grains, forming a void around it. They observed the vortices along with shock structures in the fore-wake and wake of the obstacle. Charan et al.\cite{harishmdturbulencevortices2016} carried out molecular dynamics simulations to study the flow (at low Reynolds number) past an obstacle in the visco-elastic complex medium. The emergence of vortex street structures behind the obstacle was observed when flow became laminar to turbulent. Such vortex street structures behind obstacles in the flowing dusty plasma are even possible at low Raynold numbers. Another many-body simulation study \cite{turbulence2dlayer2021} and numerical simulation \cite{vortexmovementprl20214} in a 2D dusty plasma confirms the occurrence of turbulence in dusty plasma at low-to-medium Reynolds numbers ($R< 100)$. Various studies have also explored that dust-acoustic waves in dusty plasma can trigger turbulence \cite{Zhdanonwaveturbulence2015}. Pramanik et al.\cite{pramanikwaveturbulence2003} studied the nonlinear or self-excited dust-acoustic wave turbulent state of dusty plasma at low dissipation losses. The experimental work by Tsai et al.\cite{dawturbulencedustyexp2012,liniwaveturbulenceprl2019} and Zhdanov et al.\cite{Zhdanonwaveturbulence2015} also confirmed the self-excited dust-acoustic wave turbulence in dusty plasma with decreasing the dissipation losses. In 2015, an experimental study by Tsai et al.\cite{liniacousticvortices2014} reported the self-excited fluctuating acoustic vortex pair in defect-mediated dust-acoustic wave turbulence. These acoustic vortex pair has helical waveforms oppositely winding around the hole filaments in a given plane \cite{liniacousticvortices2014}. Po-Cheng Lin and Lin I \cite{acousticvorticesinteraction2018} extended this work to explore the dust-acoustic turbulence experimentally. Authors used a multidimensional empirical mode decomposition technique to decompose the dust-acoustic wave turbulence and observed the multi-scale interacting dust acoustic vortices around the wormlike cores\cite{acousticvorticesinteraction2018}. Thus, it can be stated that vortices in dusty plasma are intricately connected to turbulence. 
%%%%%%%%%%%%%%%%%%%%%%%%
\section{Summary and Future perspective of vortex flow studies} \label{sec:secVII}
In summary, extensive studies on the dynamical vortex and coherent structures in dusty plasma have been performed
either by an individual researcher or a research group in the last 30 years. The observed dust vortices are divided into externally driven and instabilities driven depending on the free
energy source to drive the vortex flow. The external electromagnetic perturbation can modify the dynamics
of background species (electrons, ions, and neutrals), forming dust vortices through a momentum transfer mechanism between the plasma species/neutrals and charged dust grains. The previous studies show that inhomogeneity in background plasma parameters (in the plasma volume or near the surfaces) can give rise to the dust charge gradient and ions
drag gradient along a particular direction. The dust charge gradient and ion drag gradient along with
nonparallel E-field component/gravity are assumed to be the free energy sources to drive the vortex
motion against the dissipation losses of dust components. It has also been demonstrated that RT instability, K-H instability, and dust-acoustic instability can provide free energy to the dust plasma system. As a result of
these instabilities, various kinds of vortex/coherent structures are observed in the dust grain medium. The stabilizing
rate of such dusty plasma instabilities is decided by various factors such as viscosity of the medium,
compressibility of the dust-grain medium, Coulomb coupling strength amongst charged dust grains, background E-field, frictional coupling between dust-neutral, etc. Some interesting results, such as the emergence of shear waves, inertial waves, the origin of tripolar vortices, etc., during the evolution of vortices in dusty plasma medium, were also discussed. In the presence of an external magnetic field, $\vec{E} \times \vec{B}$ drifted motion of background species (electron, ions, neutrals)
or gradient-driven forces (ion drag and dust charge gradient) are responsible for the vortex formation.
The role of boundary conditions on the flowing dusty plasma around an object or the void (virtual object) in
dusty plasma is highlighted. The value of the Raynolds number in such flowing dust grain medium
determines the flow characteristics (laminar or turbulent) in the boundary layer and the vortex
formation in the wake region of the object. The past study also demonstrates that there is a connection between the
turbulent state of dust grain medium and vortices.\\\\
%%%%%%%%%%%%%%%%%%%%%%%
It has been discussed that the study of vortices/coherent structures in dusty plasma with or without B-field is an active research topic. A wide spectrum of work has been performed, but still, there are a lot of open problems. The author tries to highlight some future research problems in this review article. 
\begin{itemize}
    \item An enormous analytical and computer simulation research work on the K-H instability-driven vortices in the dusty plasma has been performed in the last few years. However, more attention is required to experimentally explore the uni-directional shear-induced vortices, the role of an external B-field on the growth and evolution of K-H instability-driven vortices, etc. 
    \item The experimental realization (verification) of analytical and simulation results of RT instability in an unstable equilibrium of two different dust grain mediums in the presence of gravity is still challenging. How do we create a dusty plasma system with a density variation opposite gravity? The role of external perturbation induced by the magnetic field, electric field, neutral flow, etc. on the RT instability and its long-time evolution.\\
    \item The experimental realization of the tripolar vortices in unmagnetized or magnetized dusty plasma needs a lot of attention.  
    \item The external B-field changes the dust charge as well as potential distribution around the floating/biased object in the dusty plasma. Therefore, a detailed study of vortices behind the biased/floating object (in the wake region) in flowing dusty plasma is needed.
    \item  The role of boundary conditions (potential boundary) on the flow characteristics of bounded and flowing dusty plasma in the absence or presence of B-field needs to be studied.
    \item The study of hexagonal vortices in simulation as well as in experiments needs to be explored.
\end{itemize}
%%%%%%%%%%%
%%%%%%%%%%%%%%%%%%%%%%%%
\section{ACKNOWLEDGMENTS}
The Author is thankful to Publishers (APS, AIP, Springer Nature, Cambridge University Press, IOP sciences) for allowing him to reuse the published figures in this new review article.   
%%%%%%%%%%
\section{Author Declarations}
\subsection{Conflict of Interest}
The author has no conflicts to disclose.
\subsection{Author Contributions}
Author contribution is not applicable to this article. All roles were played by Mangilal Choudhary in this study.   
%%%%%%%%%
\section{Data Availability}
Data sharing is not applicable to this article as no new data were created or analyzed in this study.
%%%%%%%%%%%%%%
\section{References}
\bibliography{reviewreferences}
\end{document}